\documentclass[prb,showpacs,twocolumn,superscriptaddress,amsmath,amssymb,floatfix]{revtex4-1}
\usepackage{graphicx}
\usepackage{bm}
\usepackage{bbm}
\usepackage{color}
\usepackage{comment}
\usepackage{footnote}
\usepackage{epstopdf}
\usepackage{color, colortbl}
\usepackage[table]{xcolor}
\usepackage{wrapfig}
\usepackage{hyperref}
\usepackage{upgreek}
\usepackage{appendix}

\begin{document}
\title{Bulk odd-frequency pairing in the superconducting Su-Schrieffer-Heeger model}
\author{Shun Tamura}
\author{Sho Nakosai}
\affiliation{Department of Applied Physics, Nagoya University, Nagoya 464--8603, Japan}
\author{Annica M. Black-Schaffer}
\affiliation{Department of Physics and Astronomy, Uppsala University, Box 516, S-751 20 Uppsala, Sweden}
 \author{Yukio Tanaka} 
\affiliation{Department of Applied Physics, Nagoya University, Nagoya 464--8603, Japan}
\author{Jorge Cayao}
\affiliation{Department of Physics and Astronomy, Uppsala University, Box 516, S-751 20 Uppsala, Sweden}

\date{\today} 
\begin{abstract}
The Su-Schrieffer-Heeger model describes fermions that hop on a one-dimensional chain with staggered hopping amplitude, where the unit cell contains two sites, or two sublattices. In this work we consider  the Su-Schrieffer-Heeger model with superconducting pairing and show that the sublattice index acts as  an additional quantum number in the classification of Cooper pairs, giving rise to  inter- and intra-sublattice  odd-frequency pair correlations in the bulk. 
Interestingly, this system behaves as a two band superconductor  where the bulk  odd-frequency  correlations  depend solely on the intrinsic staggering properties of the model. 
In general, odd-frequency correlations coexist with even-frequency correlations in both the trivial and topological phases, with comparable and even larger odd-frequency amplitudes at  the topological phase transition points at low frequencies, due to the  closing of the energy gap at these points.  
Furthermore, we also discuss how bulk odd-frequency amplitudes are correlated with  pseudo-gaps in the density of states and also with a charge density wave that  appears due to the chemical potential imbalance between sublattices.

\end{abstract}
\maketitle
\section{Introduction}%
\label{introduction}
Superconducting properties are highly determined by the symmetry of the Cooper pair wave function, or pair amplitude. Due to the fermionic nature of electrons,  Fermi-Dirac statistics imposes antisymmetry  on the pair amplitude  under the total exchange of all quantum numbers, which can include spin, orbital, band, spatial and time coordinates, etc. The antisymmetry condition allows for electrons to pair at equal times but, interestingly, also permits pairing at different times, giving rise to temporally non-local  pair amplitudes, which can be odd under the exchange of time coordinates, or equivalently odd in frequency ($\omega$).\cite{bere74,RevModPhys.77.1321,doi:10.1143/JPSJ.81.011013,RevModPhys.91.045005,cayao2019odd} 

Odd-frequency (odd-$\omega$) pairing can emerge as a bulk or induced phenomena and it has been shown that the main ingredient in both cases relies on breaking the system symmetries.\cite{RevModPhys.77.1321,doi:10.1143/JPSJ.81.011013,RevModPhys.91.045005,cayao2019odd}  For instance, it has been shown that odd-$\omega$ correlations can be induced in normal-superconductor junctions,\cite{PhysRevB.76.054522,Eschrig2007,Golubov2011x,doi:10.1143/JPSJ.81.011013,PhysRevLett.99.037005,PhysRevB.86.174514,Higashitani2009,PhysRevB.99.184501,PhysRevB.98.075425,PhysRevB.93.024509,cayao2019odd}
 where their emergence   occurs because the spatial parity of Cooper pairs is broken at the junction interface.
Moreover, under the presence of a spin field in such junctions, e.g.~from a magnetic field or spin-orbit coupling,  spins  get mixed, which then gives rise to even more exotic spin triplet odd-$\omega$ correlations.\cite{RevModPhys.77.1321,doi:10.1143/JPSJ.81.011013,RevModPhys.91.045005,cayao2019odd} This has allowed for an understanding of a number of exotic phenomena:  long-range proximity effect\cite{RevModPhys.77.1321,Samokhvalov_2016} and paramagnetic Meissner effect\cite{doi:10.1143/JPSJ.66.2556,PhysRevB.72.140503,PhysRevLett.106.246601,PhysRevB.89.184508,PhysRevB.91.214510} in superconductor-ferromagnet junctions, anomalous proximity-effect in spin-triplet superconductor junctions,~\cite{Kashiwaya_2000,PhysRevLett.98.037003,
PhysRevB.70.012507,PhysRevB.71.094513,PhysRevLett.96.097007,PhysRevLett.99.067005} Majorana zero modes\cite{PhysRevLett.72.1526,doi:10.1143/JPSJ.81.011013,PhysRevB.87.104513,PhysRevB.87.220506,PhysRevB.91.054518,PhysRevB.91.174511,PhysRevB.92.100507,PhysRevB.92.121404,PhysRevB.95.174516,10.1093/ptep/ptw094,PhysRevB.95.184506,PhysRevB.96.155426,Tanaka2018,PhysRevB.97.075408,PhysRevB.97.134523,PhysRevB.100.115433,cayao2019odd,dushko20} and surface impedance in topological superconductors.\cite{PhysRevLett.107.087001} 
Nowadays,  induced odd-$\omega$ pairing in junctions is well established even  experimentally,\cite{bernardo15,PhysRevX.5.041021} which reflects the vast activity made towards the understanding of this induced effect.

Odd-$\omega$ pairing can also appear as a bulk effect in  superconductors, without the need of interfaces.  This occurs in systems with multiple degrees of freedom such as multiband superconductors,\cite{PhysRevB.88.104514,PhysRevB.92.094517,PhysRevB.92.224508,doi:10.1002/andp.201900298}, double quantum dots,\cite{PhysRevB.90.220501,PhysRevB.93.201402} and double nanowires,\cite{10.1093/ptep/ptw094,PhysRevB.100.024512}  
  where the band, dot, or wire indices, respectively, allow for a more broadened family of Cooper pair symmetries where odd-$\omega$  correlations can then emerge without the need of interfaces as in junctions. In particular,  
  in  multiband superconductors it has been   shown that  odd-$\omega$ pairing can be correlated with observable signatures such as gaps in the density of states (DOS) at higher energies\cite{PhysRevB.92.094517} and the Kerr effect.\cite{PhysRevLett.119.087001,PhysRevB.97.064505}  The higher energy gaps  result from the hybridization of normal bands,\cite{PhysRevB.88.104514} which can be seen as an intrinsic symmetry breaking phenomenon, unlike the extrinsic interfaces present in junctions.

Another interesting route to bulk odd-$\omega$ correlations is through inherent staggered properties. For instance, it has been  shown that in  buckled quantum spin Hall insulators bulk odd-$\omega$ correlations appear due to a staggered order parameter but   no sign of  higher energy gaps in the DOS was reported. \cite{PhysRevB.96.174509} Even more interestingly, it has  recently been shown in nanowires with Rashba spin-orbit coupling  that intrinsic staggered hopping and spin-orbit coupling can induce a topological superconducting phase that completely repels the topological phase of the uniform nanowire.\cite{kobialka2019dimerization} 
Since these nanowires  represent  one of the most investigated platforms for one-dimensional topological superconductivity,\cite{Aguadoreview17,LutchynReview08,tkachov19review} the reported staggered-induced topological  phase is within experimental reach and its fundamental understanding is, therefore, important. Generally, this can be achieved by a proper investigation of the emergent superconducting pair correlations,  which has not been carried out so far; odd-$\omega$ correlations are expected to play an important role due to the topological nature of this exotic staggered phase, as has been shown for other topological systems.\cite{doi:10.1143/JPSJ.81.011013,PhysRevB.87.104513,PhysRevB.87.220506,PhysRevB.91.054518,PhysRevB.91.174511,PhysRevB.92.100507,PhysRevB.92.121404,PhysRevB.95.174516,10.1093/ptep/ptw094,PhysRevB.95.184506,PhysRevB.96.155426,Tanaka2018,PhysRevB.97.075408,PhysRevB.97.134523,PhysRevB.100.115433,cayao2019odd,dushko20} 
As a consequence, a general exploration of systems with intrinsic staggered properties is timely, can reveal the emergence of exotic bulk superconducting properties, and add fundamental understanding to these effects.

\begin{figure}[!t]
   \centering
   \includegraphics[width=6cm]{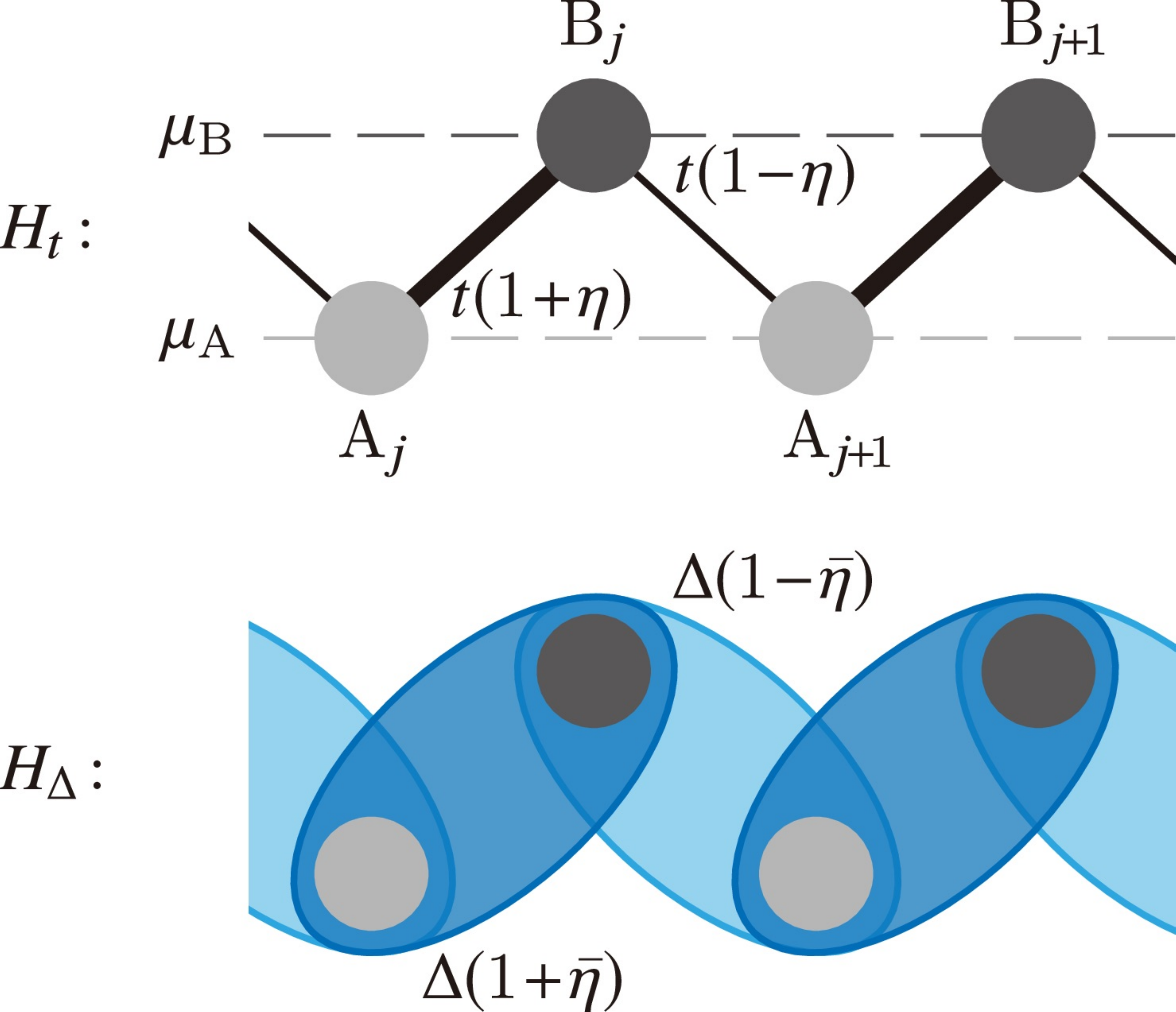}
   \caption{%
      Schematic illustration of the SC SSH model. Top: SSH model with chemical potential imbalance between sublattices A and B. Bottom: representation of the considered pair potential, which includes pairing of states of the same cell (A-B inter-sublattice pairing) and pairing of states in neighboring cells (A-B inter-sublattice pairing).
   }%
   \label{fig:schematic}
\end{figure}

In this work we consider the possibly simplest staggered superconducting system: the Su-Schrieffer-Heeger (SSH) model with superconducting (SC) correlations,  as  shown in Fig.\,\ref{fig:schematic}, and study the emergence of bulk odd-$\omega$ pair correlations. The SSH model, initially proposed in the context of the polymer polyacetylene,\cite{PhysRevLett.42.1698,PhysRevB.22.2099}  describes fermions that hop on a one-dimensional chain with staggered hopping amplitude and also exhibits a topological phase. Here, the unit cell contains  two sites, denoted as A and B and here referred to as the A and B sublattices. The spin  is absent in the SSH model, which can then be interpreted as a spinless or  spin-polarized system.\cite{asboth2016short}   
We consider  the simplest superconducting pairing in the SC SSH model, including intra and intercell nearest-neighbor
pairing.\cite{PhysRevB.90.014505,PhysRevB.89.115430} This system has been shown to  exhibit a one-dimensional topological superconducting phase with Majorana zero modes (MZMs).\cite{PhysRevB.90.014505,PhysRevB.89.115430} MZMs appear exponentially localized at the ends and their pair amplitudes have been shown to exhibit  odd-$\omega$ symmetry.\cite{doi:10.1143/JPSJ.81.011013,PhysRevB.87.104513,PhysRevB.87.220506,PhysRevB.91.054518,PhysRevB.91.174511,PhysRevB.92.100507,PhysRevB.92.121404,PhysRevB.95.174516,10.1093/ptep/ptw094,PhysRevB.95.184506,PhysRevB.96.155426,Tanaka2018,PhysRevB.97.075408,PhysRevB.97.134523,PhysRevB.100.115433,cayao2019odd,dushko20} MZMs have attracted an enormous amount of attention due to their potential use for building topological protected qubits.\cite{Aguadoreview17,LutchynReview08,tkachov19review,zhangreview}
Interesting, the  SC SSH model captures the main intrinsic staggered properties of systems that are under active investigation,\cite{Aguadoreview17,LutchynReview08,tkachov19review,zhangreview}  such as in buckled quantum spin Hall insulators\cite{PhysRevB.96.174509} and in nanowires with Rashba spin-orbit coupling.\cite{kobialka2019dimerization}  Therefore, the study of pair correlations in the SC SSH model is expected to  provide fundamental understanding of superconducting correlations in these systems.

We demonstrate that, under general conditions,  there is a coexistence of bulk even- and odd-$\omega$ correlations in the trivial and topological phases of the SC SSH model, where both amplitudes develop  intra-  and inter-sublattice components. We have performed this analysis both in momentum (infinite system) and real space (finite system), where in the latter case the  bulk pair amplitudes are probed far from the edges.
We find that the bulk odd-$\omega$ correlations emerge due to intrinsic properties of the SC SSH model. In fact, while the bulk intra-sublattice  odd-$\omega$ terms only depend on the staggered hopping and pair potential, the bulk inter-sublattice  odd-$\omega$ component necessitates both a finite chemical potential sublattice imbalance  and finite pair potential.  
Although the bulk odd-$\omega$ terms exhibit small values almost everywhere, interestingly, its low-$\omega$ components   are enhanced at the topological phase transition due to closing of the energy gap,  acquiring  values comparable  to those of the  even-$\omega$ amplitudes.
 Our findings of bulk odd-$\omega$ pair amplitudes in the SC SSH model are consistent with Fermi-Dirac statistics, where the additional degree of freedom, offered by the sublattice indices A and B,   extends the classification of superconducting correlations. This is similar to what has been reported in multiband superconductors,\cite{PhysRevB.88.104514,PhysRevB.92.094517,PhysRevB.92.224508} double quantum dots,\cite{PhysRevB.90.220501,PhysRevB.93.201402} and double nanowires.\cite{10.1093/ptep/ptw094,PhysRevB.100.024512}  
Moreover, when the system is of finite length, we find that, at the edges, the low frequency odd-$\omega$ components in the topological phase develop  larger values than the even-$\omega$ terms  due to  MZMs,\cite{doi:10.1143/JPSJ.81.011013,PhysRevB.87.104513,PhysRevB.87.220506,PhysRevB.91.054518,PhysRevB.91.174511,PhysRevB.92.100507,PhysRevB.92.121404,PhysRevB.95.174516,10.1093/ptep/ptw094,PhysRevB.95.184506,PhysRevB.96.155426,Tanaka2018,PhysRevB.97.075408,PhysRevB.97.134523,PhysRevB.100.115433,cayao2019odd,dushko20} an effect we explicitly associate with the topological bulk invariant through the so-called spectral edge boundary correspondence.\cite{PhysRevB.99.184512,PhysRevB.100.174512}

We also explore possible measurable signatures of the obtained bulk odd-$\omega$ correlations in the SC SSH model. We find that the DOS develops higher energy pseudo-gaps, whose existence correlates with the emergence of odd-$\omega$  amplitudes, similar to what has been reported for certain multiband superconductors.\cite{PhysRevB.88.104514} 
Furthermore, we discover that  the inter-sublattice odd-$\omega$ amplitudes are  finite when each sublattice has a different chemical potential, an effect that also gives rise to  a finite charge density wave (CDW),\cite{PhysRevLett.86.1857,PhysRevLett.108.136803,ezawa2013topological,szumniak2019hinge} whose non-zero value then automatically signals the emergence of   inter-sublattice odd-$\omega$ correlations. Lastly, we find that the emergence of MZMs, e.g.~as a zero energy peak in local density of states (LDOS), is a clear signal of large inter- and intra-sublattice odd-$\omega$ amplitudes. 
 Odd-$\omega$ correlations  can thus provide fundamental understanding of specific features in the DOS and CDW state in the SC SSH model.

The remainder of this article is organized as follows. In Sec.\,\ref{Model} we present the model and outline the employed method. In Sec.\,\ref{Results1} we present our results demonstrating the emergence of bulk odd-$\omega$ pairing. 
We then, in the same section, study the relation to superconducting fitness\cite{PhysRevB.98.024501,PhysRevB.100.104501} and show that it captures the conditions for odd-$\omega$ pairing. 
In  Sec.~\ref{Results2}, we  discuss odd-$\omega$ pairing in the finite length SC SSH model. In Sec.~\ref{experiments}, we explore some of the possible signatures of  bulk odd-$\omega$ correlations.
 Finally, we present our conclusions in Sec.\,\ref{conclusions}.  


\section{Model and method}%
\label{Model}
We consider  the  SC SSH model with intra- and inter-sublattice superconducting pairing  as schematically shown in Fig.~\ref{fig:schematic} and modeled  in real space by $H=H_{\rm t}+H_{\Delta}$, with
 \begin{equation}
    \label{modelSSHSC}
    \begin{split}
       H_{\rm t}&=
       -\sum_{j}\Big[\mu_{\rm A}c^{\dagger}_{{\rm A},j}c_{{\rm A},j}+\mu_{\rm B}c^{\dagger}_{{\rm B},j}c_{{\rm B},j} \Big] \\
          &-t\sum_{j}\Big[
          (1+\eta)c^{\dagger}_{{\rm B},j}c_{{\rm A},j}
          +
          (1-\eta)c^{\dagger}_{{\rm A},j+1}c_{{\rm B},j}+{\rm H.c.}
       \Big]\,,\\
  H_{\Delta}&=
          \Delta\sum_{j}
          \Big[
             (1+\bar{\eta})c^{\dagger}_{{\rm B},j}c^{\dagger}_{{\rm A},j}
             +
             (1-\bar{\eta})c^{\dagger}_{{\rm A},j+1}c^{\dagger}_{{\rm B},j}+{\rm H.c.}
          \Big]\,,
       \end{split}
   \end{equation}
where $c_{\mathrm{A(B)},j}$ $(c_{\mathrm{A(B)},j}^\dagger)$ is 
an annihilation (creation) operator on $j$-th cell and A(B) sublattice. 
Note that Eq.\,(\ref{modelSSHSC}) is in real space and models the finite length SC SSH system.
The first and second lines in Eq.\,(\ref{modelSSHSC}) correspond to the usual SSH model in the normal state but where each sublattice A and B have a different chemical potential $\mu_{\mathrm{A,B}}$, thus giving rise to a CDW.\cite{ezawa2013topological} The hopping between sites/sublattices is represented by $t>0$, with $\eta$ being the staggering parameter in the hopping term.
The third term is the superconducting pairing,\cite{PhysRevB.90.014505,PhysRevB.89.115430} where pairing  occurs between A and B sublattices both in the same cell $j$  and also between neighboring cells, with $\Delta$ being the pair potential and $\bar{\eta}$  models the staggering in the  pair potential.   
This represents the simplest possible pairing in the SC SSH model and has been shown to be relevant in previous studies.\cite{PhysRevB.90.014505,PhysRevB.89.115430} 
Here the staggering in the hopping and pairing, caused by non-zero $\eta$ and $\bar{\eta}$, assumed real without loss of generality,  is an intrinsic property of the SC SSH model. 
Moreover, the imbalance of chemical potentials between the A and B sublattices in Eq.\,(\ref{modelSSHSC})  gives rise to a CDW gap\cite{ezawa2013topological} of $\mu_{\rm A}-\mu_{\rm B}$ in the normal energy versus momentum dispersion (see below)  which will be relevant when looking at signatures of odd-$\omega$ pairing in experimental observables. 

In order to visualize the emergence of bulk odd-$\omega$ pair correlations, we perform the analysis in momentum space, as  is common for bulk properties. We first Fourier transform Eq.~(\ref{modelSSHSC}) into momentum space, which in the Nambu basis
$\Psi_{k}=(c_{{\rm A},k},c_{{\rm B},k},c^{\dagger}_{{\rm A},-k},c^{\dagger}_{{\rm B},-k})$, then reads,
\begin{equation}
\label{HSCSSHk}
H(k)=
\begin{pmatrix}
H_{0}(k)&\Delta(k)\\
\Delta^{\dagger}(k)&-H_{0}^{*}(-k)
\end{pmatrix}
\end{equation}
where 
\begin{equation}
\label{H0}
H_{0}(k)=
\begin{pmatrix}
-\mu_{\rm A}&T_{k}\\
T^{*}_{k}&-\mu_{\rm B}
\end{pmatrix}\,,\quad
\Delta(k)=\begin{pmatrix}
0&\Delta_{k}\\
-\Delta^{*}_{k}&0
\end{pmatrix}
\end{equation}
and 
\begin{equation}
\label{TDelta}
\begin{split}
   T_{k}&=-t[(1+\eta)+(1-\eta){\rm e}^{-ika}]\,,\\
\Delta_{k}&=-\Delta[(1+\bar{\eta})-(1-\bar{\eta}){\rm e}^{-ika}]\,,\\
\end{split}
\end{equation}
correspond to the staggered momentum-dependent hopping  and staggered-momentum dependent pairing, respectively. From Eqs.\,(\ref{TDelta}), we write  $\Delta_{k}=-\Delta\{[(1+\bar{\eta})-(1-\bar{\eta}){\rm cos}(ka)]+i[(1-\bar{\eta}){\rm sin}(ka)]\}$, where the first (second) term in square brackets  is even (odd) in momentum $k$, and can be seen as  $s$- and $p$-wave components, respectively.

\begin{figure}[!t]
   \centering
   \includegraphics[width=8.5cm]{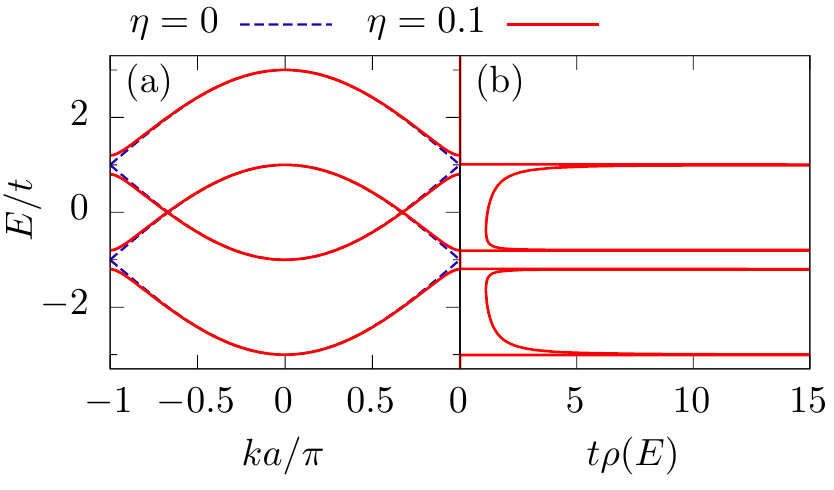}
   \caption{%
      (a) Energy versus momentum dispersion  $\pm E_\pm$ for $\Delta=0$, $\bar{\eta}=0$ and $\mu_\mathrm{A}=\mu_\mathrm{B}=t$ for $\eta =0$  (blue) and $\eta=0.1$ (red).
      (b) DOS projected on the particle space for $\eta=0.1$.
        }%
   \label{fig:eigen_value_normal}
\end{figure}
\subsection{Energy versus momentum dispersion}%
\label{sec:energy_dispersion}

Before going further we inspect the energy versus momentum dispersion of the SC SSH model in order to obtain an understanding of its intrinsic properties, which we later use when exploring detection schemes of odd-$\omega$ correlations. By diagonalizing Eq.~(\ref{HSCSSHk}) we obtain  four energy bands given by $\pm E_{\pm}$, with 
\begin{align}
E_{\pm}(k)
   =&
   \frac{1}{\sqrt{2}}\sqrt{%
      \mu_\mathrm{A}^2
   +\mu_\mathrm{B}^2+
   2
   |T_k|^2
   +2|\Delta_k|^2
\pm\sqrt{Y_k}}\,,
   \label{eq:energy_disp}
\end{align}
where
\begin{align}
   Y_k
&=
   4
  \left[
     {|T_k\Delta^*_k
      +
   T^*_k\Delta_k|}^2
   + |\Delta_k|^2
   {\left(
   \mu_\mathrm{A}
   -
   \mu_\mathrm{B}
   \right)}^2
\right]
   \nonumber\\
   &
    +
   4|T_k|^2
   {\left(
   \mu_\mathrm{A}
   +
   \mu_\mathrm{B}
   \right)}^2
   +
   {\left(
   \mu_\mathrm{A}^{2}
   -
   \mu_\mathrm{B}^{2}
   \right)}^2\,.
   \label{eq:Yk}
\end{align}
In the normal state, i.e.~$\Delta=0$, the energy bands are depicted in Fig.~\ref{fig:eigen_value_normal}(a). The electron (hole)  energy bands $E_{\pm}$ ($-E_{\pm}$) cross at higher energies,  $E_{+}=E_{-}$, at $k=\pm \pi/a$ when $\eta=0$, $T_{k}=0$, and $\mu_{\rm A}=\mu_{\rm B}$,  as seen in blue curves  in Fig.~\ref{fig:eigen_value_normal}(a). 
Interestingly, when either $T_{k}\neq0$  or $\mu_{\rm A}\neq\mu_{\rm B}$, the normal electron (hole) bands hybridize and open   higher energy gaps of size $E_{+}-E_{-}$, at $k=\pm \pi/a$, see red curves  in Fig.~\ref{fig:eigen_value_normal}(a). In what follows we refer to these higher energy gaps as hybridization gaps.
In particular, if $T_{k}=0$, the hybridization gap  is given by $E_{+}-E_{-}=(\mu_{\rm B}-\mu_{\rm A})$, which corresponds to the CDW gap\cite{ezawa2013topological}  due to the chemical  potential imbalance between sublattices. 
On the other hand, when  $\mu_{\rm A}=\mu_{\rm B}$,  the  hybridization gap is given by  $E_{+}-E_{-}=2|T_{k}|$ for $\mu_{\rm A}>|T_{k}|$ or $E_{+}-E_{-}=2|\mu_{\rm A}|$ for $\mu_{\rm A}<|T_{k}|$. Remarkably, the hybridization gaps seen in the energy bands (red curves in (a)) manifest in the electronic DOS as higher energy gaps as well, as  seen in Fig.~\ref{fig:eigen_value_normal}(b).

A finite value of the pair potential, $\Delta\neq0$, opens a gap at the Fermi points (zero-energy crossings between electron and hole bands)  in the normal spectrum, as shown in Fig.\,\ref{fig:eigen_value}(a-c) for different values of the chemical potentials.  Interestingly, the higher energy hybridization gaps observed in the normal DOS are also partly captured in the DOS at finite $\Delta$, as seen in Fig.~\ref{fig:dos_bulk} for different values of the chemical potentials. However, since the higher energy gaps in the DOS at finite $\Delta$ are not fully developed, we refer to them as pseudo-gaps.  
Further details of the energy bands and DOS at finite $\Delta$ are discussed in Subsection \ref{sec:winding_num}.

\subsection{Winding number}%
\label{sec:winding_num}
The SC SSH model also exhibits interesting topological superconducting properties due to its intrinsic features. In what follows we  analyze the topological phases of the SC SSH model given by Eq.~(\ref{HSCSSHk}) for a non-zero pair potential $\Delta$.
In order to characterize the different topological phases in the SC SSH model we calculate the winding number associated with $H(k)$ in Eq.~(\ref{HSCSSHk}).  $H(k)$  has chiral symmetry and anticommutes  with a chiral operator $\Gamma=\sigma_0\tau_1$, with the winding number~\cite{PhysRevB.83.224511} defined as
\begin{equation}
\label{winding}
   W
   =
   \frac{i}{4\pi}\int_{-\pi/a}^{\pi/a} \mathrm{d} k\:
   \mathrm{tr}
   \left[
      \Gamma H^{-1}(k) \partial_k H(k)
   \right]\,,
\end{equation}
where $\sigma_{0}$ and $\tau_{1}$ are Pauli matrices in  sublattice and particle-hole spaces, respectively. 

\begin{figure}[!t]
   \centering
   \includegraphics[width=8.5cm]{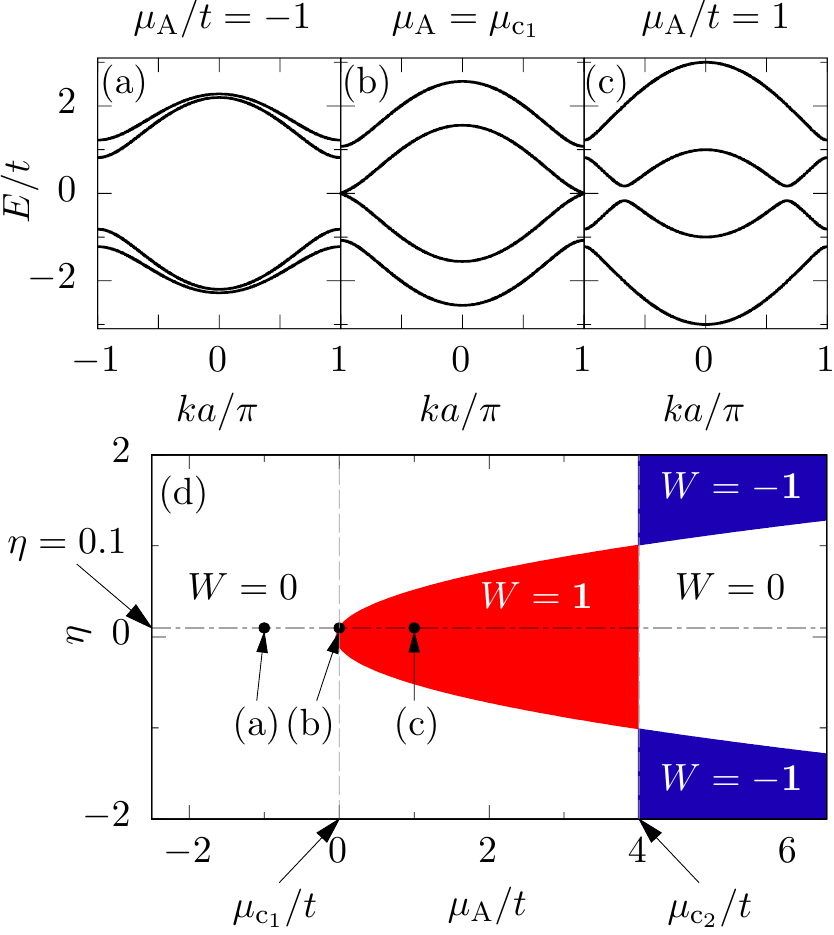}
   \caption{Energy versus momentum dispersion for (a) trivial ($\mu_\mathrm{A}/t=-1$),
      (b) topological phase transition ($\mu_\mathrm{A}=\mu_\mathrm{c_1}$), and (c) topological regimes ($\mu_\mathrm{A}/t=1$).     
      (d) The winding number $W$ as a function of $\mu_\mathrm{A}$ and $\eta$, with red and blue regions corresponding to  topological phases $W=\pm1$, respectively.     
      Parameters:  $\Delta/t=0.1$, $\mu_\mathrm{B}/t=1$ and $\bar{\eta}=0.2$. 
   In (a-c)   $\eta=0.1$ is depicted as filled circles in (d).
   }%
   \label{fig:eigen_value}
\end{figure}

In Fig.~\ref{fig:eigen_value}(d) we present the winding number $W$   as a function of $\eta$ and $\mu_\mathrm{A}$ for $\Delta/t=0.1$, $\mu_{\rm B}/t=1$,  $\bar{\eta}=0.2$.   As can be seen, depending on the system parameters, $W$ takes the values $0$ and $\pm1$, which corresponds to the trivial and topological phases, respectively.  For instance, the topological phase with $W=1$ occurs for $\mu_{c_{1}}<\mu_{\rm A}<\mu_{c_{2}}$ for a given $\eta$, where $\mu_{c_{1,2}}$ correspond to critical values that determine the phase boundary or topological phase transition (TPT).
For a finite length  SC SSH model in the topological phases with $W=\pm1$, the system  hosts MZMs at its end points,\cite{PhysRevB.90.014505,PhysRevB.89.115430} as expected from the bulk-boundary correspondence.

   \begin{figure}[!t]
   \centering
   \includegraphics[width=8.5cm]{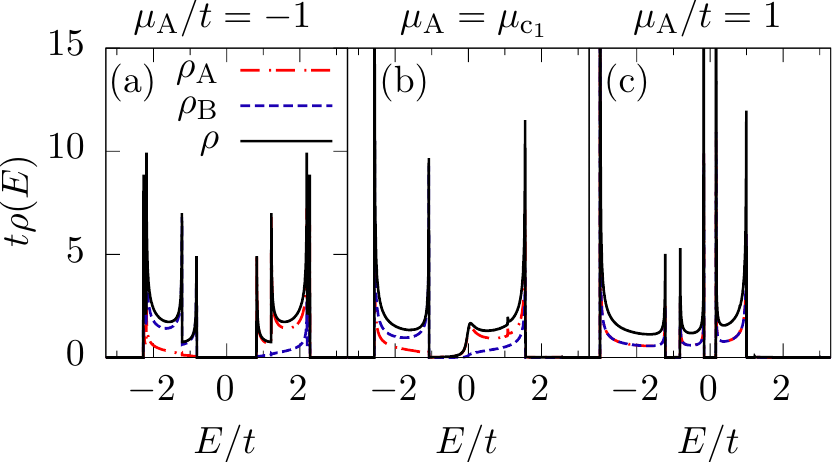}
   \caption{Particle DOS in the bulk of the SC SSH model as a function of $E$ for (a) trivial ($\mu_\mathrm{A}/t=-1$), (b) topological phase transition ($\mu_\mathrm{A}=\mu_\mathrm{c_1}$), and (c) topological ($\mu_\mathrm{A}/t=1$) regimes.
 Parameters: $\Delta/t=0.1$, $\mu_\mathrm{B}/t=1$, $\eta=0.1$ and $\bar{\eta}=0.2$.
   }%
   \label{fig:dos_bulk}
\end{figure}

Further insights to the different phases of the SC SSH model are obtained from the fact that topological phases can be distinguished by looking at the regimes where the band gap closes. 
To see this, in Fig.~\ref{fig:eigen_value} (a-c) we show the energy dispersion at non-zero $\Delta$,  given by Eq.~(\ref{eq:energy_disp}), where  we fix the parameters as in (d) and take $\eta=0.1$ but  vary $\mu_{\rm A}$ such that it captures the different phases shown in (d) at the filled black circles.  
In the trivial phase (a), the energy gap at the Fermi energy ($E=0$) is larger than $\Delta$ since it opens even for the normal state with $\Delta=0$. At the TPT in (b), the energy gap closes and marks the topological transition point. The condition for the closing of the gap can be obtained from $E_{-}(k)=0$ leading to the following two conditions:
\begin{align}
   0
   =&
   -\mu_\mathrm{A}\mu_\mathrm{B}
   +
   2
   \left[
     t^2(1+\eta^2)
      -
      \Delta^2(1+\bar{\eta}^2)
   \right]
   \nonumber\\
   &+   2 \left[
      t^2(1-\eta^2)
     +
      \Delta^2(1-\bar{\eta}^2)
   \right]
   \cos (ka),
   \label{eq:phase_boundary1}
   \\
   0
   =&
   (1-\eta \bar{\eta})\sin (ka)\,.
   \label{eq:phase_boundary2}
\end{align}
which have to be simultaneously satisfied.  In the topological phase (c) the energy gap is approximately given by $\Delta$ since the SC gap now opens at  the Fermi points. 

The pseudo-gaps in the  particle DOS at finite $\Delta$, introduced in the previous subsection, in the different phases of the SC SSH model is visualized in Fig.~\ref{fig:dos_bulk}. In the trivial phase, Fig.~\ref{fig:dos_bulk}(a), the pseudo-gaps appear at $E/t\sim\pm1$ due to the opening of the hybridization gap in the energy bands at $ka=\pm\pi$ in Fig.~\ref{fig:eigen_value}(a). 
At the topological phase transition (b), the lowest energy gap closes at the Fermi energy and there is a pseudo-gap at $E/t\sim-0.5$ due to the opening of a hybridization gap at $ka=\pm\pi$ reported in Fig.~\ref{fig:eigen_value}(b). In the topological phase (c), a small pseudo  gap at $E/t\sim -1$ emerges as seen in Fig.~\ref{fig:dos_bulk}(c), which is again the result of a band hybridization in the energy bands, presented in  Fig.~\ref{fig:eigen_value}(c). Therefore, we conclude that band hybridization gives rise to pseudo-gaps at higher energies in the DOS of the SC SSH model, which is also similar to what occurs in certain two band superconductors.\cite{PhysRevB.88.104514,PhysRevB.92.094517,PhysRevB.92.224508,doi:10.1002/andp.201900298}

\subsection{Green's function method}%
\label{sec:greens_function}
In this work we aim at investigating the pair correlations in the SC SSH model. 
For this purpose we calculate the full system Green's function, 
\begin{equation}
\label{greens}
{\bf G}(i\omega)=\begin{pmatrix}
G_{0}&F\\
\check{F}&\check{G}_{0}
\end{pmatrix}={(i\omega -H)}^{-1}
\end{equation}
where $H$ is given by Eq.\,(\ref{HSCSSHk}) in momentum space or by Eq.\,(\ref{modelSSHSC}) in real space. The matrix form of ${\bf G}$   represents the Nambu space due to the electron-hole symmetry of $H$, where $G_{0}$ and  $F$ correspond to the regular and anomalous Green's functions. Due to the basis of $H$, the structure of the Green's function components is
\begin{equation}
\label{FG}
   G_{0}(i\omega)=
   \begin{pmatrix}
      G_{\rm AA}&G_{\rm AB}\\
      G_{\rm BA}&G_{\rm BB}
   \end{pmatrix},\,
   F(i\omega)=
   \begin{pmatrix}
      F_{\rm AA}&F_{\rm AB}\\
      F_{\rm BA}&F_{\rm BB}
   \end{pmatrix},
\end{equation}
where $F_{\rm AA,BB}$ and $F_{\rm AB,BA}$ correspond to intra-sublattice and inter-sublattice pair correlations, originated from the sublattice indices A and B. 
Moreover, each element of the anomalous term $F_{ab}(i\omega)$, represents a pair amplitude with momentum $k$ or  spatial coordinates in real space, frequency $\omega$, and sublattice  dependence $a,b=$A, B. The sublattice index, therefore, extends the classification of Cooper pairs in this system, where the symmetries of $F_{\rm AA, BB,AB}$ determine the symmetries of the superconducting correlations and  play a crucial role for the emergence of bulk odd-$\omega$ correlations in the SC SSH model, as discussed next.  This view is further supported by previous studies in multiband superconductors,\cite{PhysRevB.88.104514,PhysRevB.92.094517,PhysRevB.92.224508} double quantum dots,\cite{PhysRevB.93.201402} and double nanowires\cite{10.1093/ptep/ptw094} where the band, dot, and wire indices, respectively, played the role of the sublattice index discussed here in the SC SSH model. We stress that the sublattice degree of freedom represents  an intrinsic property of the SC SSH model considered in Eq.\,(\ref{modelSSHSC}) and, therefore, does not depend on external considerations such as e.g., interfaces in junctions.

From the Green's functions in Eq.\,(\ref{greens}) we can also calculate experimental observables that are important for the characterization of superconducting correlations. For instance, the normal Green's function $G_0$ allows the calculation of the DOS $ \rho(E)
   =
   \rho_\mathrm{A}(E)
   +
   \rho_\mathrm{B}(E)$, where
\begin{equation}
   \label{eq:rho_A_B}
  \begin{split}
     \rho_\mathrm{A(B)}(E)
   &=
   -\frac{1}{\pi}\mathrm{Im}\{G_\mathrm{AA(BB)}(i\omega=E+i\delta)\},
\end{split}
\end{equation}
represents the DOS at sublattice A(B), with $E$ being the real energy. We assume $\delta/\Delta=10^{-5}$  throughout this work  and verify that it is sufficiently small  to not alter our calculations.


\section{Bulk pair correlations}%
\label{Results1} 
In this part we investigate the bulk  pair correlations in the SC SSH model.
By using Eq.\,(\ref{greens}) we obtain the pair amplitudes,
\begin{equation}
   \label{Pairamplitudes}
   \begin{split}
      F_{\rm AA}(i\omega,k)&=\frac{\mu_{\rm B}(\Delta^{*}_{k}T_{k}-\Delta_{k}T^{*}_{k})-i\omega(\Delta^{*}_{k}T_{k}+\Delta_{k}T_{k}^{*})}{P(i\omega,k)}\\
      F_{\rm BB}(i\omega,k)&=\frac{\mu_{\rm A}(\Delta^{*}_{k}T_{k}-\Delta_{k}T^{*}_{k})+i\omega(\Delta^{*}_{k}T_{k}+\Delta_{k}T_{k}^{*})}{P(i\omega,k)}\\
      F_{\rm AB}(i\omega,k)&=
      \frac{1}{P(i\omega,k)}\Big[i\omega \Delta_{k}(\mu_{\rm B}-\mu_{\rm A})\\
                           &+\Delta^{*}_{k}(T_{k}^{2}-\Delta_{k}^{2})-\Delta_{k}(\omega^{2}+\mu_{\rm A}\mu_{\rm B})\Big]\\
      F_{\rm BA}(i\omega,k)&=
      \frac{1}{P(i\omega,k)}\Big[i\omega \Delta_{k}^{*}(\mu_{\rm B}-\mu_{\rm A})\\
                           &-\Delta_{k}[{(T_{k}^{*})}^{2}-{(\Delta_{k}^{*})}^{2}]+\Delta_{k}^{*}(\omega^{2}+\mu_{\rm A}\mu_{\rm B})\Big]\,,
   \end{split}
\end{equation}
where $P(i\omega,k)$ is a frequency and momentum dependent real polynomial in even powers of $\omega$ and $k$ whose explicit expression is not important for our discussion here but is for completeness given in Appendix~\ref{sec:deno_green_app}. 

The main  observation from Eqs.\,(\ref{Pairamplitudes}) is that there is a linear frequency dependent term in both the intra-   $F_{\rm AA,BB}$ (second term) and inter-sublattice $F_{\rm AB, BA}$ (first term) pair amplitudes. These linear frequency components represent the odd-$\omega$ pair correlations. 
The odd-$\omega$ amplitudes describe a pair of electrons at different times, which vanish at equal times,  thus reflecting that
  odd-$\omega$ pairing is  an intrinsically dynamical phenomenon\cite{RevModPhys.77.1321,doi:10.1143/JPSJ.81.011013,RevModPhys.91.045005,cayao2019odd}  that differs from the standard  even-$\omega$ pairing.
Further understanding of the pair correlations in the SC SSH model is given by analyzing the symmetries of the pair amplitudes in Eqs.~(\ref{Pairamplitudes}) in terms of its frequency, momentum, and sublattice indices. 

In the case of   intra-sublattice amplitudes, given by the first two  relations in Eqs.~(\ref{Pairamplitudes}),  we notice that  $F_{\rm AA, BB}$ have two clear components, with the first term being even in $\omega$ and the second term being odd in $\omega$.  
The even-$\omega$ term, $(\Delta^{*}_{k}T_{k}-\Delta_{k}T^{*}_{k})=4it\Delta(1-\eta\bar{\eta})\sin(ka)$, is clearly odd in momentum $k$ where we have used the expressions for $\Delta_{k}$ and $T_{k}$ given by Eqs.~(\ref{TDelta}). On the other hand,  the odd-$\omega$ term, $\Delta^{*}_{k}T_{k}+\Delta_{k}T_{k}^{*}=
4t\Delta[\eta+\bar{\eta}+(\bar{\eta}-\eta)\cos(ka)]$, is even in momentum $k$.  Note that the  intra-sublattice odd-$\omega$ amplitudes $F_{\rm AA, BB}$  are finite when either $\bar{\eta}$ or $\eta$  is non-zero ($t$ and $\Delta$ are always assumed to be non-zero), and thus a consequence of the intrinsic sublattice symmetry breaking in the SC SSH model\cite{PhysRevB.90.014505,PhysRevB.89.115430}  but also predicted to appear in other systems such as in buckled quantum spin Hall insulators,\cite{PhysRevB.96.174509} or in nanowires with Rashba spin-orbit coupling.\cite{kobialka2019dimerization}   For $\eta=\bar{\eta}=0$ the intra-sublattice odd-$\omega$ amplitudes vanish. At $\eta\bar{\eta}=1$, however, the even-$\omega$ component is zero, leaving only finite intra-sublattice odd-$\omega$ correlations. Furthermore, the intra-sublattice pairing is automatically even (E) under the exchange of sublattice indices. This, together with the discussion above, implies that the  intra-sublattice pair amplitudes  given in Eqs.~(\ref{Pairamplitudes}) can be classified as EEO and OEE symmetry classes in the frequency-sublattice-momentum nomenclature.

\begin{figure*}[!t]
   \centering
   \includegraphics[width=15cm]{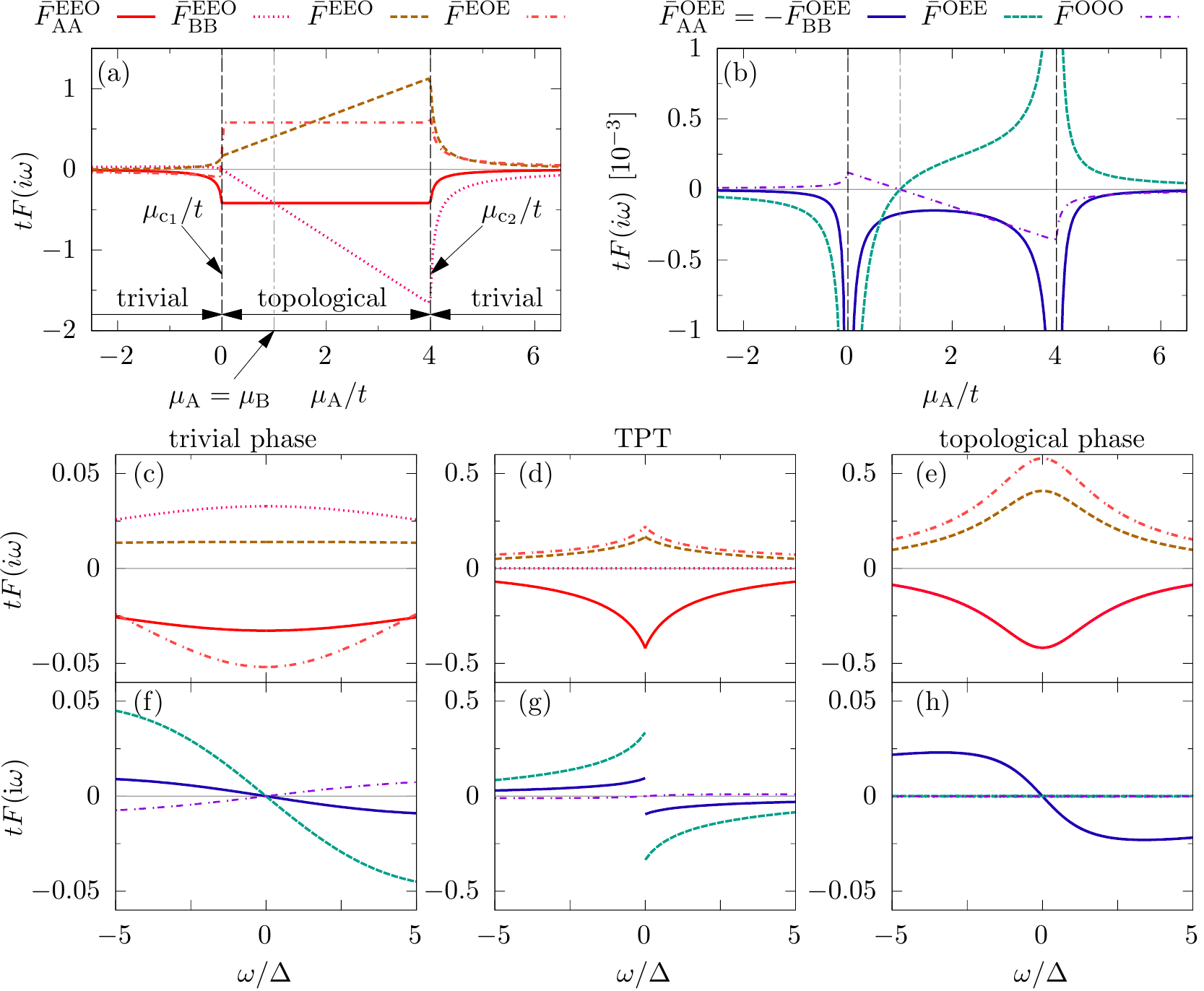}
   \caption{%
      Bulk pair amplitudes as a function of  $\mu_\mathrm{A}$ at $\omega/\Delta=10^{-2}$ (a,b) and  $\omega$  (c-h) for $\Delta/t=0.1$,    $\mu_\mathrm{B}/t=1$, $\eta=0.1$, and $\bar{\eta}=0.2$. Vertical dashed lines in (a,b) indicate the TPTs $\mu_{\rm c_{1,2}}$, where $\mu_\mathrm{c_1}<\mu_\mathrm{A}<\mu_\mathrm{c_2}$ is the topological phase. Dot-dashed vertical lines in (b) mark the points where inter-sublattice OEE and OOO vanish. Panels (c-e) and (f-h) represent the frequency dependent even- and odd-$\omega$ pair amplitudes, respectively, with  $\mu_{\rm A}$ as in Fig.\,\ref{fig:eigen_value}, namely, $\mu_\mathrm{A}/t=-1$, $\mu_\mathrm{A}=\mu_{\rm c_{1}}$, and $\mu_\mathrm{A}=t$ for the trivial, topological phase transition, and topological regimes, respectively.
      }%
      \label{fig:agreen_bulk}
\end{figure*}

For the inter-sublattice pair symmetries we proceed as in the previous paragraph, but before continuing we write the even and odd combinations  under the exchange of A,B: $F_{\pm}(i\omega,k)=[F_{\rm AB}(i\omega,k)\pm F_{\rm BA}(i\omega,k)]/2$.  
Then, by a close inspection, we obtain that each, $F_{+}$ and $F_{-}$, has  even- and odd-$\omega$ components whose symmetries are classified as OOO and OEE in the frequency-sublattice-momentum nomenclature. Their explicit expressions are:
 $F_{+}^{\rm OEE}(i\omega,k)=-2i\omega(\mu_{\mathrm{B}}-\mu_{\mathrm{A}}) \Delta[1+\bar{\eta}+(\bar{\eta}-1)\cos(ka)]/P(i\omega,k)$ 
and $F_{-}^{\rm OOO}(i\omega,k)=2\omega(\mu_{\mathrm{B}}-\mu_{\mathrm{A}})\Delta(\bar{\eta}-1)\sin(ka)/P(i\omega,k)$, with an evident even and odd momentum dependence, respectively, in order to fulfill Fermi-Dirac statistics. To obtain  the expressions for OOO and OEE amplitudes we have used the expression for $\Delta_{k}$ given in Eq.\,(\ref{TDelta}). Importantly, the non-zero value of  inter-sublattice odd-$\omega$ terms,  $F_{+}^{\rm OEE}$ and $F_{-}^{\rm OOO}$, is conditioned to the finite chemical potential imbalance between  sublattices A and B, whose size corresponds to the CDW gap.\cite{ezawa2013topological} In addition, also non-zero values of $\bar{\eta}$, $\Delta$ are needed. 

Following a similar inspection as discussed above, we find that the inter-sublattice pair amplitudes $F_{\pm}$ also have even-$\omega$ components, corresponding to $F_{+}^{\rm EEO}$ and $F_{-}^{\rm EOE}$, and noticeable from the second and third terms in the last two expressions of Eqs.\,(\ref{Pairamplitudes}). These even-$\omega$ amplitudes represents a more standard type of superconducting pairing,\cite{RevModPhys.77.1321,doi:10.1143/JPSJ.81.011013,RevModPhys.91.045005,cayao2019odd}  which can account for  pairing of electrons at equal times, unlike the odd-$\omega$ correlations.

In order to visualize the discussion made in the previous few paragraphs,  we define 
 the ``$s$-wave'', which is even in $k$, and the ``$p$-wave'', which is odd in $k$, components of the anomalous Green's function, as
   \begin{equation}
    \label{eq:F_s}
\begin{split}     
      \bar{F}_{ab}^s(i\omega)
      &=
      \frac{a}{2\pi}
      \int_{-\pi/a}^{\pi/a} \mathrm{d}k
      F_{ab}(i\omega,k),
      \\
      \bar{F}_{ab}^p(i\omega)
      &=
      \frac{a}{2\pi}
      \int_{-\pi/a}^{\pi/a} \mathrm{d}k
      F_{ab}(i\omega,k)\sin (ka).
      \end{split}
   \end{equation}
    where $a$ and $b$ denote sublattices A or B and $ F_{ab}$ correspond to the pair amplitudes given by Eqs.~(\ref{Pairamplitudes}). Then, in order to account for the symmetries discussed above, we also define
   \begin{equation}
   \label{despairs}
   \begin{split}
      \bar{F}_\mathrm{AA(BB)}^\mathrm{OEE}(i\omega)
      =&
      \mathrm{Im}
      \bar{F}_\mathrm{AA(BB)}^s(i\omega),
      \\
      \bar{F}^\mathrm{OEE}(i\omega)
      =&
      \mathrm{Im}
      \bar{F}_\mathrm{AB+BA}^s(i\omega),
      \\
      \bar{F}^\mathrm{EOE}(i\omega)
      =&
      \mathrm{Re}
      \bar{F}_\mathrm{AB-BA}^s(i\omega),
      \\
      \bar{F}_\mathrm{AA(BB)}^\mathrm{EEO}(i\omega)
      =&
      \mathrm{Im}
      \bar{F}_\mathrm{AA(BB)}^p(i\omega),
      \\
      \bar{F}^\mathrm{EEO}(i\omega)
      =&
      \mathrm{Im}
      \bar{F}_\mathrm{AB+BA}^p(i\omega),
      \\
      \bar{F}^\mathrm{OOO}(i\omega)
      =&
      \mathrm{Re}
      \bar{F}_\mathrm{AB-BA}^p(i\omega),
      \end{split}
   \end{equation}
   with   
   $\bar{F}_{\mathrm{AB}\pm\mathrm{BA}}^{s(p)}(i\omega)
   =
[
      \bar{F}_{\mathrm{AB}}^{s(p)}(i\omega)
      \pm
      \bar{F}_{\mathrm{BA}}^{s(p)}(i\omega)
]/2$. In writing Eqs.\,(\ref{despairs}) we have utilized that 
  $\bar{F}_{\mathrm{AA,BB,AB+BA}}^{s(p)}$ and $\bar{F}_{\mathrm{AB-BA}}^{s(p)}$  solely exhibit imaginary and real values, respectively, as shown in Appendix \ref{sec:green_derivation_app}. Note also that, by using Eqs.~(\ref{Pairamplitudes}), we can show that $\bar{F}_{\mathrm{AA}}^{\mathrm{OEE}}(i\omega)=-\bar{F}_{\mathrm{BB}}^{\mathrm{OEE}}(i\omega)$, and, therefore, it is only necessary to investigate and plot one of them.

In Figs.~\ref{fig:agreen_bulk} we  plot the  pair amplitudes given by Eqs.~(\ref{despairs}) as a function of the chemical potential $\mu_\mathrm{A}$ at $\omega/\Delta=10^{-2}$ (a,b) and frequency $\omega$ (c-h), where in the former we also  detect the different topological phases. 
The even-$\omega$ components  develop generally larger values in the topological phase than in the trivial phase, as seen in  (a). These even-$\omega$ correlations usually exhibit also larger amplitudes than the odd-$\omega$  terms in both the trivial and topological phases, as seen by comparing (a) and (b). 
However,  at the TPTs  ($\mu_\mathrm{c_1}$ and $\mu_\mathrm{c_2}$ marked with dashed vertical lines) both the intra- and inter-sublattice amplitudes $\bar{F}_\mathrm{AA(BB)}^\mathrm{OEE}$ and  $\bar{F}^\mathrm{OEE}$ have divergent profiles with comparable or even larger values than their even-$\omega$ counterparts.
The inter-sublattice OOO component also exhibits its maximum value at TPTs but with an overall smaller amplitude than the even-$\omega$ terms.
The divergent values at the TPTs occur because the denominator of the pair amplitudes, $P(i\omega,k)$ in Eq.~(\ref{Pairamplitudes}), has a singularity at the energy gap closing.  Furthermore, at $\mu_\mathrm{A}=\mu_\mathrm{B}$ (vertical dot-dashed grey line in Fig.\,\ref{fig:agreen_bulk}(a,b)), the inter-sublattice amplitudes $\bar{F}^\mathrm{OEE}$  and $\bar{F}^\mathrm{OOO}$  vanish, leaving only non-zero intra-sublattice odd-$\omega$ correlations with  OEE symmetry. This can be also directly seen in Eqs.~(\ref{Pairamplitudes}).

Next we investigate the  frequency dependence of the pair amplitudes in Figs.~\ref{fig:agreen_bulk}(c--h). As expected, $\bar{F}_{\mathrm{AA(BB)}}^{\mathrm{EEO}}$, $\bar{F}^{\mathrm{EEO}}$, and  $\bar{F}^{\mathrm{EOE}}$ exhibit an  even  frequency dependence   with their maximum  absolute value  occurring at $\omega=0$ in the trivial, TPT, and topological phases,  as seen in (c-e) .  Likewise, the odd-$\omega$ components, $\bar{F}_{\mathrm{AA(BB)}}^{\mathrm{OEE}}$, $\bar{F}^{\mathrm{OEE}}$, and $\bar{F}^{\mathrm{OOO}}$, display the expected odd frequency dependence. 
However, unlike all the even-$\omega$ terms,  the intra-sublattice OEE and inter-sublattice OEE amplitudes develop a  discontinuous profile with a maximum magnitude around $\omega=0$ only at the TPT,  which can be understood as a result of  the energy gap closing at the TPT; the inter-sublattice OOO component at the TPT is instead very small and with smooth profile across $\omega=0$. In the trivial (f) and topological phases (h) all odd-$\omega$ components have a smooth and approximate linear frequency dependence at low frequencies.

We thus find that intra- and inter-sublattice odd-$\omega$ amplitudes, given by Eqs.~(\ref{Pairamplitudes}), do not vanish when $T_{k}\neq0$ and $\mu_{\rm A}\neq\mu_{\rm B}$, respectively, provided $\Delta_{k}\neq0$. Interestingly, the presence of these bulk odd-$\omega$ correlations can be correlated with experimental observables as well. In fact, in Subsection \ref{sec:winding_num} we show that either $T_{k}\neq0$ or $\mu_{\rm A}\neq\mu_{\rm B}$ induce a hybridization of the normal electron (hole) bands in the SC SSH model  which  then give rise to pseudo gaps in the DOS, as seen in Fig.~\ref{fig:dos_bulk}. Therefore, this allows us to conclude that the pseudo gaps in the DOS  indicates the presence of either intra- or inter-sublattice odd-$\omega$ correlations, as the conditions for these two effects to emerge coincide. We point out that these features are similar to what occurs in certain two band superconductors.\cite{PhysRevB.88.104514,PhysRevB.92.094517,PhysRevB.92.224508,doi:10.1002/andp.201900298}

To summarize this part, we stress that  the SC SSH model hosts bulk dynamic odd-$\omega$ correlations, due to its intrinsic properties, namely, staggered hopping, staggered pair potential, and chemical potential imbalance. 
  While even-$\omega$ correlations develop larger values than the odd-$\omega$ amplitudes in the trivial and topological phases,  at the TPTs the odd-$\omega$ amplitudes  exhibit a  discontinuous profile with  maximum magnitude whose  values  are comparable or even larger than even-$\omega$ amplitudes.
Our findings, therefore, suggest that both even- and odd-$\omega$ correlations must be considered when studying superconductivity in the SC SSH model.

\subsection{Superconducting fitness}
\label{scfitness}
All the  conditions that we have identified so far for the emergence of finite odd-$\omega$ correlations  can be elegantly obtained by calculating the superconducting fitness,\cite{PhysRevB.98.024501,PhysRevB.100.104501} $C=H_{0}(k)\Delta(k)-\Delta(k)H^{*}_{0}(-k)$. It was demonstrated in Ref.\,\onlinecite{doi:10.1002/andp.201900298} that it is possible   to determine the presence of odd-$\omega$ pairing in multiband superconductors when $C\neq0$. 
 By plugging Eqs.\,(\ref{H0}) into  $C$, we obtain the following conditions
\begin{equation}
\label{SCfitness}
\begin{split}
   (\mu_{\rm A}-\mu_{\rm B})\Delta_{k}&\neq0,\\
T_{k}\Delta^{*}_{k}+\Delta_{k}T^{*}_{k}&\neq0\,,
\end{split}
\end{equation}
where only one of these expressions needs to be satisfied in order for 
odd-$\omega$ pairing to appear.  

The finite value of the odd-$\omega$ amplitudes, obtained from Eqs.~(\ref{Pairamplitudes}), is fully consistent with the expressions derived from the superconducting fitness in Eqs.\,(\ref{SCfitness}). In fact,  the first condition in  Eqs.\,(\ref{SCfitness}), clearly reflects the need of a chemical potential imbalance, which is indeed the necessary condition for finite inter-sublattice odd-$\omega$ correlations,  $F_{\rm AB, BA}$, as can be seen in Eqs.~(\ref{Pairamplitudes}). 
Similarly, the second condition captures the non-zero intra-sublattice  odd-$\omega$ correlations, $F_{\rm AA,BB}$ in Eqs.~(\ref{Pairamplitudes}).

\section{Pair correlations in the finite size  SC SSH model}%
\label{Results2}
After showing that  bulk odd-$\omega$ correlations emerge in the SC SSH model, we next explore the odd-$\omega$ amplitudes when the system has a finite size and is modeled in real space by Eq.\,(\ref{modelSSHSC}). This is particularly motivated because it has been shown that, in the topological phase, the SC SSH model hosts MZMs,\cite{PhysRevB.90.014505,PhysRevB.89.115430} and MZMs have been predicted to enhance odd-$\omega$ amplitudes at the edges of topological superconductors,
\cite{doi:10.1143/JPSJ.81.011013,PhysRevB.87.104513,PhysRevB.87.220506,PhysRevB.91.054518,PhysRevB.91.174511,PhysRevB.92.100507,PhysRevB.92.121404,PhysRevB.95.174516,10.1093/ptep/ptw094,PhysRevB.95.184506,PhysRevB.96.155426,Tanaka2018,PhysRevB.97.075408,PhysRevB.97.134523,PhysRevB.100.115433,cayao2019odd,dushko20} although they have not yet been investigated in the context of the SC SSH model.

We use a tight-binding representation for the SC SSH Hamiltonian, given by Eq.\,(\ref{modelSSHSC}), in Nambu space with  $N$ lattice unit cells and  lattice spacing $a$  in the basis $(c_{\mathrm{A}, 1},c_{\mathrm{A}, 1}^{\dagger},c_{\mathrm{B}, 1},c_{\mathrm{B}, 1}^{\dagger},\dots,c_{\mathrm{B}, N},c_{\mathrm{B}, N}^{\dagger})$. The position coordinate is then denoted as $x=ja$ with unit cell index $j$, being $1\leq j \leq N$, and  system size  $L=100a$. We then obtain the pair amplitudes from the anomalous component of the Green's function $g_{j_{1},j_{2}}(i\omega)={(i\omega-H)}_{j_{1},j_{2}}^{-1}$, where $j_{1,2}$ corresponds to two lattice sites in the tight-binding representation. The Green's function $g$ has $4N\times 4N$ elements due to the sublattice and particle-hole symmetries.  Similarly, as for the discussion of the bulk pair amplitudes  in Sec.\,\ref{Results1}, in what follows we decompose the anomalous term of $g$ into the symmetry classes defined in Eqs.\,(\ref{despairs}), which can be written as 
\begin{align}
   f_{\mathrm{AA(BB)}}^\mathrm{OEE}(i\omega,j)
   =&
   \mathrm{Im}
   f_{\mathrm{AA(BB)}}^{\mathrm{intra}}(i\omega,j),
   \label{eq:FAAs}
   \\
   f^\mathrm{OEE}(i\omega,j) =&
   \mathrm{Im}
   f_{\mathrm{AB+BA}}^{\mathrm{intra}}(i\omega,j),
   \label{eq:FABBAs}
   \\
   f^\mathrm{EOE}(i\omega,j)
   =&
   \mathrm{Re}
   f_{\mathrm{AB-BA}}^{\mathrm{intra}}(i\omega,j),
   \label{eq:FAB_BAs}
   \\
   f_{\mathrm{AA(BB)}}^\mathrm{EEO}(i\omega,j)
   =&
   \mathrm{Im}
   f_{\mathrm{AA(BB)}}^{\mathrm{n.n.}}(i\omega,j),
   \label{eq:FAAp}
   \\
   f^\mathrm{EEO}(i\omega,j)
   =&
   \mathrm{Im}
   f_{\mathrm{AB+BA}}^{\mathrm{n.n.}}(i\omega,j),
   \label{eq:FABBAp}
   \\
   f^\mathrm{OOO}(i\omega,j)
   =&
   \mathrm{Re}
   f_{\mathrm{AB-BA}}^{\mathrm{n.n.}}(i\omega,j),
   \label{eq:FAB_BAp}
\end{align}
where
\begin{figure*}[!t]
   \centering
   \includegraphics[width=16cm]{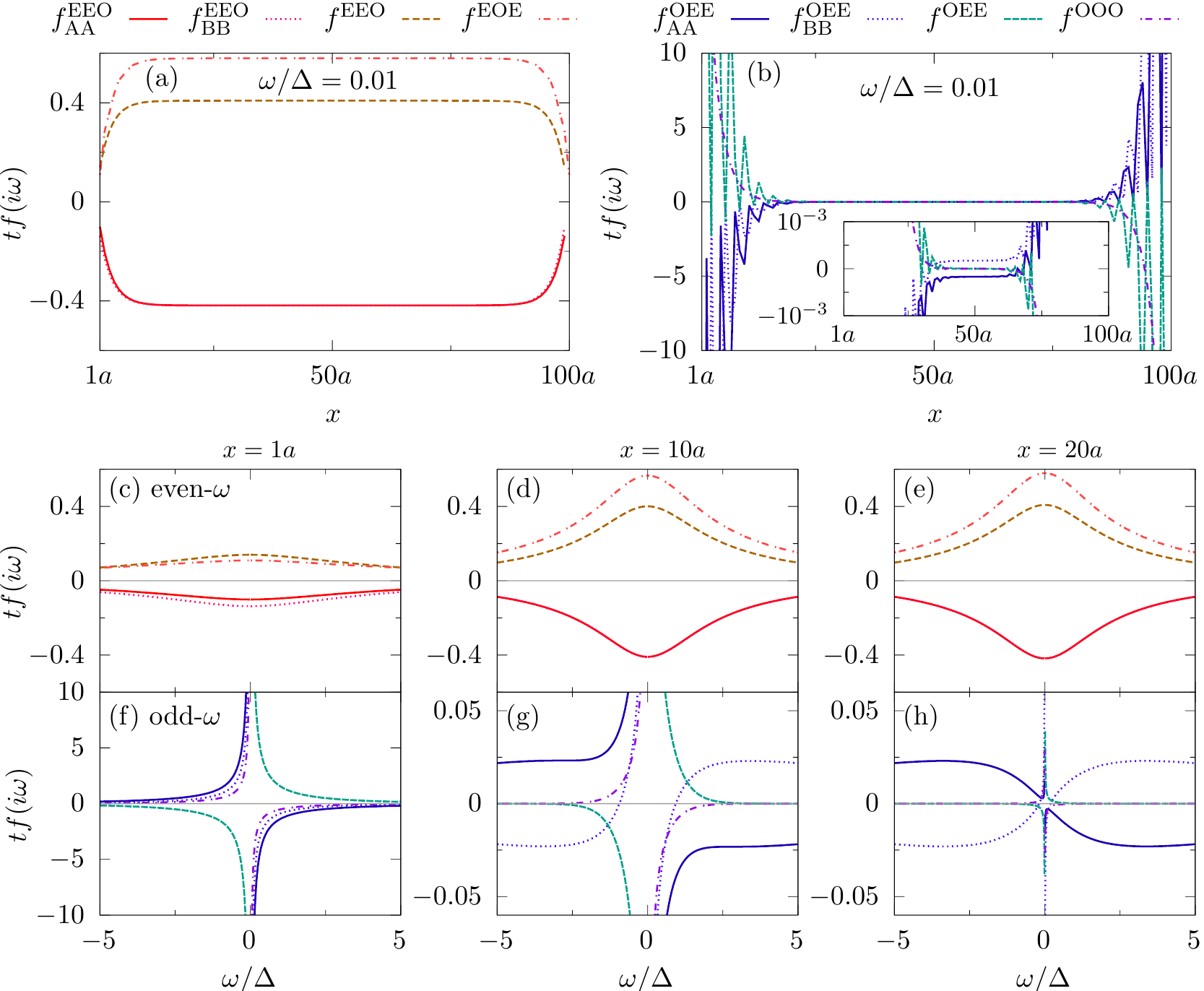}
   \caption{%
     Pair amplitudes in the topological phase of the finite length SC SSH model with $L=100a$ as a function of  (a,b) spatial coordinate $x=ja$ and (c-h) frequency. In (a,b)  $\omega/\Delta=10^{-2}$. The inset in (b) magnifies the view of the odd-$\omega$ amplitudes in the bulk.  In panels (c,f), (d,g), and (e,h) the frequency dependence of the pair amplitudes is plotted for $x=1a$, $x=10a$, and $x=20a$, respectively.   Parameters:  $\Delta/t=0.1$, $\mu_{\rm A,B}=t$, $\eta=0.1$, and $\bar{\eta}=0.2$, see Fig.\,\ref{fig:eigen_value}(c).
       }%
   \label{fig:odd_finite}
\end{figure*}
\begin{equation}
\label{eq23}
\begin{split}
   f_{\mathrm{AA}}^{\mathrm{intra}}(i\omega,j)
   &=
   g_{4j-3,4j-2}(i\omega),
   \\
   f_{\mathrm{BB}}^{\mathrm{intra}}(i\omega,j)
   &=
   g_{4j-1,4j}(i\omega),
   \\
   f_{\mathrm{AB\pm BA}}^{\mathrm{intra}}(i\omega,j)
   &=
   \frac{1}{\sqrt{2}}
   \left[
      g_{4j-3,4j}(i\omega)\pm g_{4j-1,4j-2}(i\omega)
   \right],
   \\
   f_{\mathrm{AA}}^{\mathrm{n.n.}}(i\omega,j)
   &=
   \frac{1}{2\mathrm{i}}
   \left[
      g_{4j-3,4j+2}(i\omega)-g_{4j+1,4j-2}(i\omega)
   \right],
   \\
   f_{\mathrm{BB}}^{\mathrm{n.n.}}(i\omega,j)
   &=
   \frac{1}{2\mathrm{i}}
   \left[
      g_{4j-1,8j}(i\omega)-g_{4j+3,4j}(i\omega)
   \right],
   \\
   f_{\mathrm{AB\pm BA}}^{\mathrm{n.n.}}(i\omega,j)
   &=
   \frac{1}{2\sqrt{2}\mathrm{i}}
   \left\{
      \left[
         g_{4j-3,8j}(i\omega) - g_{4j+1,4j}(i\omega)
      \right]
   \right.
   \\
    &
    \left.
       \pm
       \left[
          g_{4j-1,4j+2}(i\omega) - g_{4j+3,4j-2}(i\omega)
       \right]
    \right\},
    \end{split}
\end{equation}
correspond to intra- and nearest-neighbor (n.n) unitcell components. We have  checked that Eqs.~(\ref{eq:FAAs}), (\ref{eq:FABBAs}), (\ref{eq:FAAp}), and (\ref{eq:FABBAp}) have only imaginary parts, while  Eqs.~(\ref{eq:FAB_BAs}) and (\ref{eq:FAB_BAp}) only real, which is taken into account when discussing them next. These pair amplitudes  allow us to make a direct comparison with the results found in the bulk system in the previous section and displayed in Fig.~\ref{fig:agreen_bulk}. 
In fact, we have verified that in the bulk,  at $x=L/2$ with sufficiently large $L$, i.e.~far from both edges, the amplitudes obtained from Eqs.~(\ref{eq:FAAs})-(\ref{eq:FAB_BAp}) coincide with the amplitudes obtained with Eqs.~(\ref{despairs}), respectively. 

To proceed we concentrate on the topological phase where MZMs emerge at  both edges.  In Fig.~\ref{fig:odd_finite} we present the pair amplitudes as a function of space and frequency.  Here, the superconducting coherence length is approximately $4a$, which then indicates that the system size is much larger than the coherence length. At low frequencies, all the even-$\omega$ amplitudes show roughly constant values in the bulk away from the edges for $x\gtrsim10a$,~\cite{PhysRevB.101.024509} while at the edges  they exhibit reduced but finite values, as seen in (a).  
On the other hand, the low-frequency odd-$\omega$ components develop a huge increase near the edges, from where they decay, but do not vanish, towards the bulk of the system in an exponentially oscillatory fashion (b).   
We attribute this enhancement  to the emergence of MZMs in the topological phase, in a similar way as in other topological systems.\cite{doi:10.1143/JPSJ.81.011013,PhysRevB.87.104513,PhysRevB.87.220506,PhysRevB.91.054518,PhysRevB.91.174511,PhysRevB.92.100507,PhysRevB.92.121404,PhysRevB.95.174516,10.1093/ptep/ptw094,PhysRevB.95.184506,PhysRevB.96.155426,Tanaka2018,PhysRevB.97.075408,PhysRevB.97.134523,PhysRevB.100.115433,cayao2019odd,dushko20} 
Moreover, in order to identify the emergence of odd-$\omega$ amplitudes in the bulk  as well as compare with the results presented in the previous section, we present in the inset of Fig.~\ref{fig:odd_finite}(b), a magnified view. This probes the pair amplitudes for small $\omega/\Delta=10^{-2}$  in the center of the system, at $x\sim50a$. 
Notice that the values of the odd-$\omega$ components are almost the same as those shown in Fig.~\ref{fig:agreen_bulk}(b) with $\mu_{\mathrm{A}}/t=1$. Likewise, we have verified that the even-$\omega$ terms at $x=50a$ coincide with those  obtained using   Eqs.~(\ref{despairs}). Moreover, we stress that $f_{\mathrm{AA}}^{\mathrm{OEE}}=-f_{\mathrm{BB}}^{\mathrm{OEE}}$ is only satisfied far from both edges, in agreement with what we discuss after Eqs.\,(\ref{despairs}).

Next we turn our discussion to the frequency dependence of the pair amplitudes, presented in Figs.~\ref{fig:odd_finite}(c-h) for different locations in space. First, all the pair amplitudes have the expected even- or odd-$\omega$ dependence but the 
odd-$\omega$ correlations exhibit a dramatic change of behavior depending on if they are calculated in  the center of the system at $x=L/2$  or at the edge at $x=1a$. At the edge, all the odd-$\omega$ terms develop a drastic increase at low frequencies with a divergent profile, supporting the idea that their huge values  originate due to the presence of a MZM.\cite{doi:10.1143/JPSJ.81.011013,PhysRevB.87.104513,PhysRevB.87.220506,PhysRevB.91.054518,PhysRevB.91.174511,PhysRevB.92.100507,PhysRevB.92.121404,PhysRevB.95.174516,10.1093/ptep/ptw094,PhysRevB.95.184506,PhysRevB.96.155426,Tanaka2018,PhysRevB.97.075408,PhysRevB.97.134523,PhysRevB.100.115433,cayao2019odd,dushko20}
At $x=10a$ the odd-$\omega$ amplitudes acquire smaller values but still large variations in their frequency dependence. This is more clearly seen at  $x=20a$ where the huge values of odd-$\omega$  amplitudes around $\omega\approx0$, seen at the edge, is more narrow but still large. We have checked that deep in the bulk, $x=50a$ for this case, the odd-$\omega$ correlations in the finite size SC SSH model recover the behavior presented in Fig.~\ref{fig:agreen_bulk}(h).
On the contrary, the even-$\omega$ terms at $x=1a$ exhibits a smooth behavior as a function of frequency with non-zero values and a maximum value around low frequencies, as seen in (c). Moving towards the bulk, at  $x=10a, 20a$,  the frequency dependence of the  even-$\omega$ amplitudes is preserved, but they acquire larger values, as presented in (d,e). 


\subsection{Spectral bulk-boundary correspondence}
\label{spectralBBC}
Further understanding of the relation between   odd-$\omega$ correlations and the topological phase can be obtained from a spectral bulk-boundary correspondence (SBBC).\cite{PhysRevB.99.184512,PhysRevB.100.174512} The SBBC tells us that some components of the anomalous Green's function are related to an extended version of the winding number defined in the bulk  for chiral symmetric systems.  For the SC SSH model, which is chiral symmetric, with open boundary conditions, the amount of intra-sublattice  odd-$\omega$ components at the edge is given by  
\begin{equation}
\label{eq24}
          F_{L=Na}(i\omega)
      =
      2i
      \sum_{j=1}^{N/2}
      \left[
         f_{\mathrm{AA}}^\mathrm{OEE}(i\omega,j)
         +
         f_{\mathrm{BB}}^\mathrm{OEE}(i\omega,j)
      \right]\,.
      \end{equation}
We point out that it is due to the special form of the chiral operator for the SC SSH model, discussed in Appendix~\ref{sec:sbbc_app}, that only the intra-sublattice components OEE appear in the previous expression.  Then, following Refs.\,\onlinecite{PhysRevB.99.184512,PhysRevB.100.174512}, we  numerically  find that, at small $\omega$, the previous expression can be written as
\begin{equation}
     \label{eq:sbbc_expansion}
      \lim_{N\rightarrow\infty}
F_{L=Na}(i\omega)
      =
      W/(i\omega)+i\chi\omega+\mathcal{O}(\omega^3)\,,
   \end{equation}
where  $W$ is the winding number, given by Eq.\,(\ref{winding}), and $\chi$ is a real number. Details about the derivation can also be found in Appendix~\ref{sec:sbbc_app}.

By a simple inspection of Eq.~(\ref{eq:sbbc_expansion}), we notice that  $F_{L=Na}(i\omega)$ diverges at low frequencies in the topological phase ($W=\pm1$), while it is a linear function of $\omega$ in the trivial phase ($W=0$). This frequency dependence is in agreement with the divergent behavior in the topological phase of the low-frequency intra-sublattice pair correlations at the edge in Fig.~\ref{fig:odd_finite}(f). Interestingly, Eq.~(\ref{eq:sbbc_expansion}) indicates that it is possible to calculate some of the odd-$\omega$ correlations at the edge of the SC SSH model, which is an edge property, simply by calculating the winding number $W$ by  Eq.\,(\ref{winding}), which is a bulk property of the system. Moreover,  even though both topological phases  ($W=\pm1$) host MZMs at the system ends, their associated odd-frequency pair amplitudes  exhibit opposite sign, as seen in Eq.\,(\ref{eq:sbbc_expansion}).

\section{Further experimental consequences}%
\begin{figure}[!t]
   \centering
   \includegraphics[width=7.5cm]{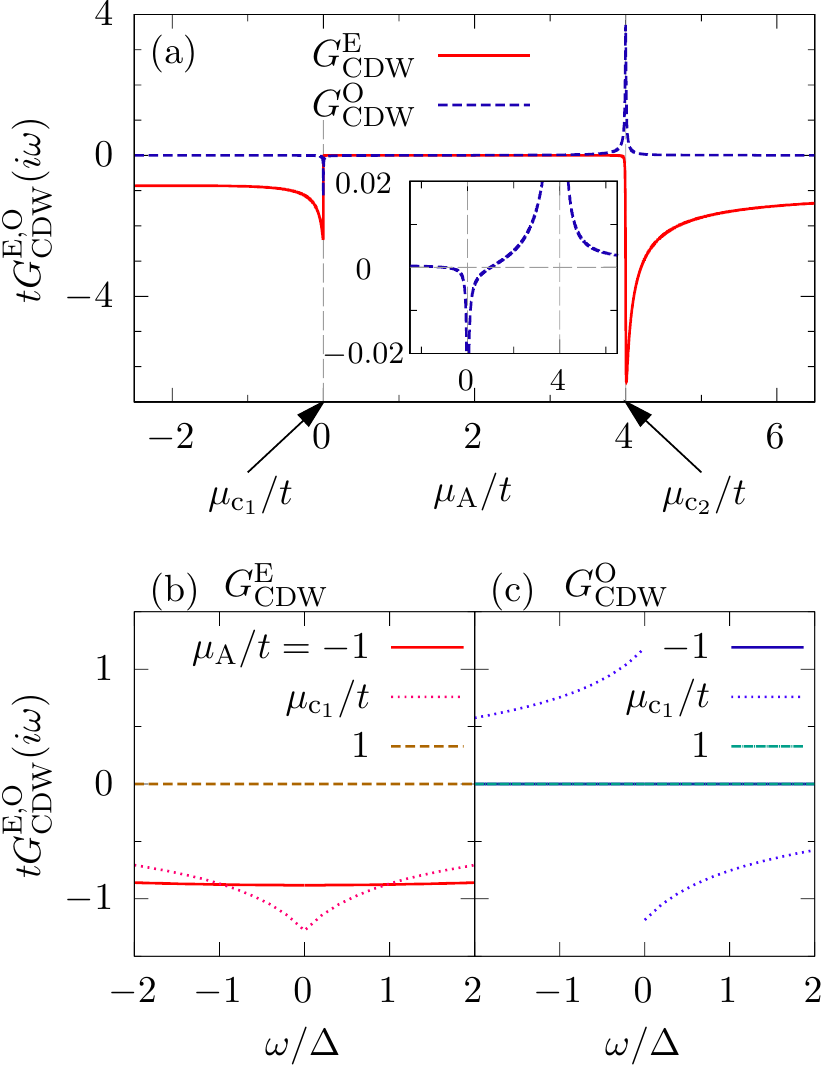}
   \caption{%
      (a) $G_\mathrm{CDW}^\mathrm{E}(i\omega)$ and $G_\mathrm{CDW}^\mathrm{O}(i\omega)$ as a function of $\mu_\mathrm{A}$ at $\omega/\Delta=10^{-2}$
      for $\Delta/t=0.1$, $\mu_\mathrm{B}/t=1$, $\eta=0.1$, and $\bar{\eta}=0.2$.
      Inset of (a) shows a magnified view of $G_{\mathrm{CDW}}^\mathrm{O}(i\omega)$.
      (b,c) Frequency dependence of $G_\mathrm{CDW}^\mathrm{E,O}(I\omega)$  for $\mu_\mathrm{A}/t=-1$, 
      $\mu_{\mathrm{c}_1}/t$, and $1$ with $\Delta/t=0.1$, $\mu_\mathrm{B}/t=1$, $\eta=0.1$, and $\bar{\eta}=0.2$.  }
      \label{fig:bulk_cdw}
\end{figure}

\label{experiments}
As we mentioned in Sec. \,\ref{Model},  the imbalance between the chemical potentials in sublattices  A and B allow for a CDW.\cite{ezawa2013topological}  
In order to characterize the CDW, which is due to a chemical potential imbalance, we define the CDW Green's function as the imbalance between regular Green's functions at sublattices A and B, namely,   $G_{\rm CDW}=G_\mathrm{AA}- G_\mathrm{BB}$, where $G_\mathrm{AA,BB}$ are obtained from Eqs.\,(\ref{greens}) and (\ref{FG}). Then, we obtain
   \begin{equation}
       \label{eq:bulk_cdw2}
   \begin{split}
    G_{\rm CDW}(i\omega,k)
      &=
      \left(
         \mu_\mathrm{A}
         -
         \mu_\mathrm{B}
      \right)
[
         {-(i\omega)}^2
         +
         \left(
            \mu_\mathrm{A}
            +
            \mu_\mathrm{B}
         \right)
         i\omega
      \\
      &+
         \left(
            |T_k|^2
            -
            |\Delta_k|^2
            -
            \mu_\mathrm{A}
            \mu_\mathrm{B}\right)
]
      /P(i\omega,k),
  \end{split}
   \end{equation}
where $P(i\omega,k)$ is  the same  real polynomial as in Eqs.\,(\ref{Pairamplitudes}) with even powers of $\omega$ and $k$. Interestingly,  Eq.\,(\ref{eq:bulk_cdw2}) is non-zero if $\mu_\mathrm{A}\neq\mu_\mathrm{B}$, the condition that also allows for finite inter-sublattice odd-$\omega$ amplitudes, as seen in Eqs.\,(\ref{Pairamplitudes}) and \,(\ref{SCfitness}). Thus,  a finite CDW is a good indicator of non-zero inter-sublattice odd-$\omega$ amplitudes. For a further interpretation and understanding, we define 
   \begin{equation}
   \begin{split}
      G_{\rm CDW}^\mathrm{E}(i\omega)
      &=
      \mathrm{Re}G_{\rm CDW}(i\omega),
      \\
      G_{\rm CDW}^\mathrm{O}(i\omega)
      &=
      \mathrm{Im}G_{\rm CDW}(i\omega),
      \\
      G_{\rm CDW}(i\omega)
      &=
      \frac{a}{2\pi}
      \int_{-\pi/a}^{\pi/a} \mathrm{d}k\:
      G_\mathrm{CDW}(i\omega,k)\,,
      \end{split}
   \end{equation}
where  $G_{\rm CDW}^\mathrm{E(O)}$  are even and odd functions of $\omega$, respectively. The behavior of these two quantities is shown in Fig.~\ref{fig:bulk_cdw}(a-c).  In panel (a) $G_\mathrm{CDW}^{\mathrm{E(O)}}$ is presented at low frequencies ($\omega/\Delta=10^{-2}$) as a function of $\mu_\mathrm{A}$. 
The main feature here is that both $G_\mathrm{CDW}^\mathrm{E,O}$  take large values at the TPTs at $\mu_\mathrm{A}=\mu_\mathrm{c_1}$ and $\mu_\mathrm{A}=\mu_\mathrm{c_2}$. As we move away from the TPTs, $G_\mathrm{CDW}^\mathrm{O}$ acquires very small values since it is an odd function of $\omega$ and we are here also at small $\omega$. On the other hand, $G_\mathrm{CDW}^\mathrm{E}$ in the trivial phase develops  smaller, but finite, values than in the topological phase. The frequency dependence of  $G_\mathrm{CDW}^\mathrm{E,O}$  is shown in panels (b,c) for $\mu_\mathrm{A}/t=-1$, $0$ and $1$, values that correspond to the trivial phase, TPT, and topological phase, respectively, also used in Fig.~\ref{fig:eigen_value}(a-c). 
These two panels  allow us to conclude that $G_\mathrm{CDW}^\mathrm{E,O}$ indeed represent even and odd-$\omega$ CDWs, respectively, where the latter was also predicted to appear in other systems.\cite{PhysRevB.64.035107,1501.07049} 

Although the condition $\mu_\mathrm{A}\neq\mu_\mathrm{B}$ simultaneously allows for finite   CDW and inter-sublattice odd-$\omega$ correlations, $F_{+}^{\mathrm{OEE}}$, it is important to disentangle which of the CDWs captures the  behavior of $F_{+}^{\mathrm{OEE}}$. The dependence on $\mu_{\rm A}$  can be observed in the inset of Fig.~\ref{fig:bulk_cdw}(a) and it has to be compared with Fig.~\ref{fig:agreen_bulk}(b), where we see that both quantities exhibit large values at the TPTs and sign change at $\mu_{\mathrm{A}}=\mu_{\mathrm{B}}$.  Moreover, the frequency dependence of the odd-$\omega$ CDW and $F_{+}^{\mathrm{OEE}}$ at the  TPT is also similar, with  a maximum and  sharp sign change at $\omega=0$, as seen by comparing Figs.~\ref{fig:bulk_cdw}(c)
and Fig.~\ref{fig:agreen_bulk}(g). We, therefore, conclude that   $F_{+}^{\mathrm{OEE}}$ and the odd-$\omega$ CDW exhibit a  similar qualitative behavior.
We can, therefore, conclude that the emergence of a CDW  signals the presence  of inter-sublattice odd-$\omega$ correlations, as both occur due to a chemical potential imbalance between sublattices in the SC SSH model. This chemical potential imbalance is due to modulations of the chemical potentials,\cite{PhysRevLett.108.136803} obtained e.g.~via electrically tunable gates.\cite{PhysRevLett.86.1857,szumniak2019hinge} 
We argue that the CDW can be measured as an imbalance of DOSs between sublattices, thus allowing a way to obtain the  inter-sublattice odd-$\omega$ correlations.

\section{Conclusions}%
\label{conclusions}
We investigated the  superconducting pair symmetries in the superconducting Su-Schrieffer-Heeger model and demonstrated that this system hosts bulk odd-frequency pair correlations in the trivial and topological phases, due to its intrinsic staggered properties. In particular, we found that the sublattice degree of freedom is  responsible for extending the classification of pair correlations, thus allowing a coexistence of inter- and intra-sublattice even- and odd-frequency amplitudes. The odd-frequency correlations  depend on the staggered hopping and chemical potential imbalance between sublattices, provided there is a non-zero pair potential,   conditions that are also consistent with the non-zero value of the superconducting fitness.

In the topological phase in finite length systems we showed that the low-frequency  odd-frequency amplitudes are enhanced at the edges of the system due to the presence of Majorana zero modes.
Furthermore, we demonstrated that the spectral bulk boundary correspondence reveals the relation between  intra-sublattice  odd-frequency  correlations and  winding number, thus extending previous studies\cite{PhysRevB.99.184512,PhysRevB.100.174512} to systems with intrinsic staggering, such as the SC SSH model. Our results  highlight odd-frequency pairing as a bulk phenomenon that goes beyond the induced effect in junctions  and does not require any external actor, such as  interfaces in junctions, but instead, this bulk effect solely relies on the intrinsic staggered nature of the system.  
 
Lastly, we  analyzed some of the possible experimental signals that can be correlated with  odd-frequency amplitudes. First, we showed that a finite staggered hopping  or chemical potential imbalance gives rise to pseudo gaps in the DOS, which then signal the  presence of  intra- or inter-sublattice odd-frequency amplitudes, respectively. 
Second, we found that the finite chemical potential imbalance between sublattices gives rise to a charge density wave whose odd-frequency component develops a qualitatively equivalent behavior as the inter-sublattice odd-frequency amplitude.  These modulations in the chemical potential can be induced via e.g.~electrically tunable gates,\cite{PhysRevLett.108.136803,PhysRevLett.86.1857,szumniak2019hinge} where the charge density wave can be measured as the imbalance of DOSs between sublattices. Finally, for a finite system length, the zero energy peak in the LDOS, which  reflects the emergence of a Majorana zero mode at the edges of the system, directly indicates the presence of large inter- and intra-sublattice odd-frequency correlations.

\section{Acknowledgements}
We thank  P. Burset and O. A. Awoga  for helpful discussions.   
S. T., S. N., and Y. T.  acknowledge the support from Grant-in-Aid for Scientific Research on Innovative Areas, Topological Material Science (Grants No. JP15H05851, No. JP15H05853, and No. JP15K21717) and Grant-in-Aid for Scientific Research B (KAKENHI Grant No. JP18H01176) from the Ministry of Education, Culture, Sports, Science, and Technology, Japan (MEXT) and JST CREST Grant No. JPMJCR16F2. Y. T. is also supported by Scientific Research A (KAKENHI Grant No. JP20H00131) and JSPS Core-to-Core program ``Oxide Superspin international network''. A. B. S. and J. C. acknowledge support from the Swedish Research Council (Vetenskapsr\aa det Grant No.~2018-03488), the Knut and Alice Wallenberg Foundation through the Wallenberg Academy Fellows program, and the European Research Council (ERC) under the European Unions Horizon 2020 Research and Innovation Programme (ERC-2017-StG-757553).

\appendix
\section{Denominator of the bulk Green's function}%
\label{sec:deno_green_app}
The denominator of the bulk Green's function in Eq.~(\ref{Pairamplitudes})  is given by
\begin{align}
   P(i\omega,k)
   =&
   {(i\omega)}^4
   -
   \left[
      2
      \left(
         {|T_k|}^2
         +
         {|\Delta_k|}^2
      \right)
      +
      \mu_\mathrm{A}^2
      +
      \mu_\mathrm{B}^2
   \right]
   {(i\omega)}^2
   \nonumber\\
    &
   +
   {|T_k^2-\Delta_k^2|}^2
   -
   2
   \left(
      {|T_k|}^2
      -
      {|\Delta_k|}^2
   \right)
   \mu_\mathrm{A}
   \mu_\mathrm{B}
   \nonumber\\
    &
   +
   {\left(
         \mu_\mathrm{A}
         \mu_\mathrm{B}
   \right)}^2\,.
   \label{eq:pomega}
\end{align}
By  simple inspection, we verify that $P(i\omega,k)$ is a real number for real  $\omega$ and an even function of $\omega$.
Moreover, it is also even in $k$ since $T_k^2$, $\Delta_k^2$ and ${|T_k^2-\Delta_k|}^2$ are all even in $k$. This can be seen by writing the following expressions,
\begin{align}
   {|T_k|}^2
   =&
   t^2
   \left[
      {(1+\eta)}^2
      +
      {(1-\eta)}^2
      +
      2
      (1-\eta^2)
      \cos (ka)
   \right],
   \\
   {|\Delta_k|}^2
   =&
   \Delta^2
   \left[
      {(1+\bar{\eta})}^2
      +
      {(1-\bar{\eta})}^2
      -
      2
      (1-\bar{\eta}^2)
      \cos (ka)
   \right],
   \\
   {|T_k^2-\Delta_k|}^2
   =&
   {|T_k|}^4
   +
   {|\Delta_k|}^4
   -
   2\mathrm{Re}{(T_k\Delta_k^*)}^2\,,
\end{align}
where we have used the definitions of $T_k$ and $\Delta_k$ given by Eqs.\,(\ref{TDelta}) in the main text.
Then, we  conclude that $P(i\omega,k)$ satisfies $P(-i\omega,-k)=P(i\omega,k)$, conditions that are used when discussing the properties of the Green's function given by Eq.~(\ref{Pairamplitudes}) in the main text.
\section{Real and imaginary parts of the bulk pair amplitudes}%
\label{sec:green_derivation_app}
In this appendix we explain why we take the real and imaginary parts of the bulk pair amplitudes given in Eqs.~(\ref{despairs}).
In Eqs.~(\ref{Pairamplitudes}), the real  and imaginary parts of $\Delta_{k}^{*}T_k$, $\Delta_k$ and $\Delta_k^*(T_k^2-\Delta_k^2)$
are given by
\begin{widetext}
\begin{equation}
\label{peeped}
\begin{split}
   \mathrm{Re}\Delta_{k}^{*}T_{k}
   &=
   2t\Delta
   \left[
   (\eta+\bar{\eta})
   +(-\eta+\bar{\eta})
   \cos(ka)
   \right],
   \\
   \mathrm{Im}\Delta_{k}^{*}T_k
   &=
   -2i\Delta t\left[1-\eta\bar{\eta}\sin (ka)\right],
   \\
   \mathrm{Re}\Delta_{k}
   &=
   -\Delta
   \left[
      (1+\bar{\eta})
      -
      (1-\bar{\eta})
      \cos(ka)
   \right],
   \\
   \mathrm{Im}\Delta_{k}
   &=
   -i\Delta
   (1-\bar{\eta})
   \sin(ka),
   \\
   \mathrm{Re}\Delta_{k}^{*}(T_{k}^{2}-\Delta_{k}^{2})
   &=
   \Delta
   \left[
      t^2(1+\eta)^{2}
      -
      \Delta^2(1+\bar{\eta})^{2}
   \right]
   [-(1+\bar{\eta})+(1-\bar{\eta})\cos(ka)]
   \\
    &+
   2
   \Delta
   \left[
      t^2(1-\eta^{2})
      +
      \Delta^2(1-\bar{\eta}^{2})
   \right]
   [-(1+\bar{\eta})\cos(ka)+(1-\bar{\eta})]
   \\
    &+
   2
   \Delta
   \left[
      t^2(1-\eta)^{2}
      -
      \Delta^2(1-\bar{\eta})^{2}
   \right]
   [-(1+\bar{\eta})\cos(2ka)+(1-\bar{\eta})\cos(ka)],
   \\
   \mathrm{Im}\Delta_{k}^{*}(T_{k}^{2}-\Delta_{k}^{2})
   &=
   i\Delta(1-\bar{\eta})
   \left[
      t^2(1+\eta)^{2}
      -
      \Delta^2(1+\bar{\eta})^{2}
   \right]
   \sin(ka)
    +2i
    \Delta(1+\bar{\eta})
   \left[
      t^2(1-\eta^{2})
      +
      \Delta^2(1-\bar{\eta}^{2})
   \right]
   \sin(ka)
   \\
    &-
   i
   \Delta
   \left[
      t^2(1-\eta)^{2}
      -
      \Delta^2(1-\bar{\eta})^{2}
   \right]
   [-(1+\bar{\eta})\sin(2ka)+(1-\bar{\eta})\sin(ka)].
   \end{split}
\end{equation}
\end{widetext}

First, by using  previous equations, $F_{\mathrm{AA(BB)}}^{s}(i\omega)$ from Eqs.\,(\ref{eq:F_s}) is given by
\begin{align}
   F_{\mathrm{AA}}^{s}(i\omega)
   =&
   -
   F_{\mathrm{BB}}^{s}(i\omega)
   \nonumber\\
   =&
   -i\frac{a}{2\pi}
   \int_{-\pi/a}^{\pi/a} \mathrm{d}k
   \frac{\omega(\Delta_{k}^{*}T_{k}+\Delta_{k}T_{k}^{*})}
   {P(i\omega,k)}\,,
\end{align}
where we can see that $F_{\mathrm{AA(BB)}}^{s}(i\omega)$ is an odd function of $\omega$, since $P(i\omega,k)$ is real, and an even function of $\omega$ and $\mathrm{Re}F_{\mathrm{AA(BB)}}^{s}(i\omega)=0$. Hence, we set $F_{\mathrm{AA(BB)}}^{\mathrm{OEE}}(i\omega)=\mathrm{Im}F_{\mathrm{AA(BB)}}^{s}(i\omega)$, which is the first equation written in Eqs.~(\ref{despairs}).

Second, similarly as before, $F^{s}_{\mathrm{AB+BA}}(i\omega)$ is given by
\begin{align}
   F_{\mathrm{AB+ BA}}^{s}(i\omega)
   =&
   \frac{i\omega a}{2\pi}
   \int_{-\pi/a}^{\pi/a} \mathrm{d}k
   \frac{(\Delta_k+\Delta_k^*)(\mu_\mathrm{B}-\mu_\mathrm{A})}
   {P(i\omega,k)}.
\end{align}
By using Eqs.\,(\ref{peeped}), we obtain that this quantity is  purely imaginary and an odd function of $\omega$. Then, we set $F^\mathrm{OEE}(i\omega)=\mathrm{Im}F_{\mathrm{AB+ BA}}^s(i\omega)$, which is the second equation in  Eqs.~(\ref{despairs}).

Third, $F^{s}_{\mathrm{AB-BA}}(i\omega)$ is given by
\begin{align}
   &F_{\mathrm{AB-BA}}^{s}(i\omega)
   \nonumber\\
   =&
   \frac{a}{\pi}
   \int_{-\pi/a}^{\pi/a} \mathrm{d}k
   \frac{\mathrm{Re}[\Delta_k^*(T_k^2-\Delta_k^2)-\Delta_k(\omega^2+\mu_\mathrm{A}\mu_\mathrm{B})]}
   {P(i\omega,k)},
\end{align}
which is purely real and an even function of $\omega$.
Then, we set $F^\mathrm{EOE}(i\omega)=\mathrm{Re}F_{\mathrm{AB+ BA}}^s(i\omega)$, giving rise to the third line  in Eqs.~(\ref{despairs}).

Fourth, $F_{\mathrm{AA(BB)}}^{p}(i\omega)$ is
\begin{equation}
\begin{split}
   &F_{\mathrm{AA(BB)}}^{p}(i\omega)=\\
   &=
   \frac{a}{2\pi}
   \int_{-\pi/a}^{\pi/a} \mathrm{d}k
   \frac{\mu_\mathrm{B(A)}(\Delta_{k}^{*}T_{k}-\Delta_{k}T_{k}^{*})\sin(ka)}
   {P(i\omega,k)}.
\end{split}
\end{equation}
Then,  $F_{\mathrm{AA(BB)}}^{p}(i\omega)$ is purely imaginary with an even dependence in $\omega$.
Therefore, we set $F_{\mathrm{AA(BB)}}^{\mathrm{EEO}}(i\omega)=\mathrm{Im}F_{\mathrm{AA(BB)}}^{p}(i\omega)$, which is the fourth expression in Eqs.~(\ref{despairs}).

Fifth, $F_{\mathrm{AB+BA}}^{p}(i\omega)$ is given by
\begin{equation}
\begin{split}
  & F_{\mathrm{AB+BA}}^{p}(i\omega)=   \frac{ia}{\pi} \times\\
   & \times \int_{-\pi/a}^{\pi/a} \mathrm{d}k
   \frac{\mathrm{Im}[\Delta_k^*(T_k^2-\Delta_k^2)-\Delta_k(\omega^2+\mu_\mathrm{A}\mu_\mathrm{B})]\sin(ka)}
   {P(i\omega,k)},
   \end{split}
\end{equation}
which is  purely imaginary and an even function of $\omega$.
Then, we set $F^\mathrm{EEO}(i\omega)=\mathrm{Im}F_{\mathrm{AB+BA}}^p(i\omega)$, leading to the fifth line in  Eqs.~(\ref{despairs}).

Sixth, for $F_{\mathrm{AB-BA}}^{p}(i\omega)$ we get
\begin{equation}
\begin{split}
   &
   F_{\mathrm{AB-BA}}^{p}(i\omega)=\\
& =
   -
   \frac{\omega a}{\pi}
   \int_{-\pi/a}^{\pi/a} \mathrm{d}k
   \frac{\mathrm{Im}(\Delta_k)(\mu_\mathrm{B}-\mu_\mathrm{A})\sin(ka)}
   {P(i\omega,k)},
   \end{split}
\end{equation}
which  is purely real and has an odd dependence on $\omega$.
Then, we denote $F^\mathrm{OOO}(i\omega)=\mathrm{Re}F_{\mathrm{AB-BA}}^p(i\omega)$, which corresponds to the last expression in Eqs.~(\ref{despairs}).

\begin{figure}[!t]
   \centering
   \includegraphics[width=8.5cm]{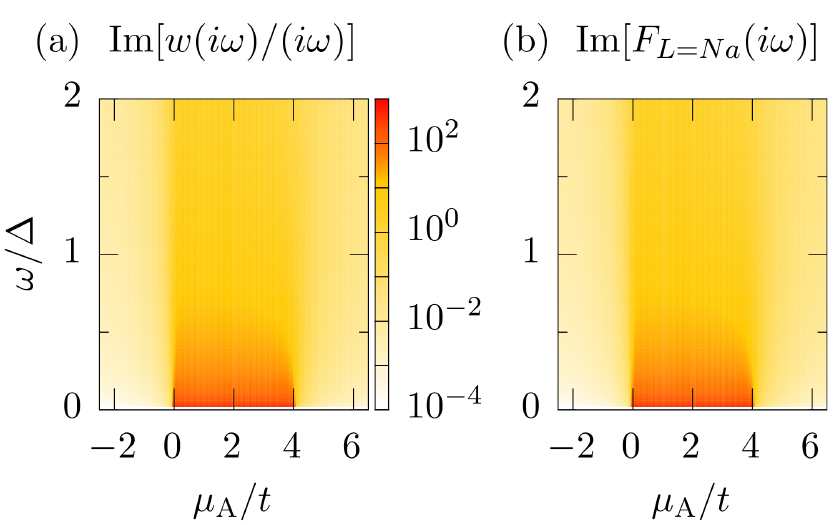}
   \caption{%
      (a) $\mathrm{Im}[w(i\omega)/(i\omega)]$, (b) $\mathrm{Im}[F_{L=Na}(I\omega)]$ with $N=1000$       are plotted as functions of $\mu_\mathrm{A}$ and $\omega$ for
      $\Delta/t=0.1$, $\mu_\mathrm{B}/t=1$, $\eta=0.1$ and $\bar{\eta}=0.2$.
      $\mathrm{Re}[w(i\omega)/(i\omega)]$ and $\mathrm{Re}[F_{L=Na}(i\omega)]$ are zero if $\omega$ is real number.
      Here the color bar is common for (a) and (b).
      }%
      \label{fig:sbbc}
\end{figure}
\section{Spectral bulk-boundary correspondence}%
\label{sec:sbbc_app}
The relation between odd-$\omega$ pair amplitudes and the topological number for a chiral symmetric system is given by the spectral bulk-boundary correspondence.~\cite{PhysRevB.99.184512,PhysRevB.100.174512} For chiral symmetric systems, following the spectral bulk-boundary correspondence, we can write\cite{PhysRevB.99.184512,PhysRevB.100.174512}
\begin{align}
   w(i\omega)/(i\omega)=\lim_{N\rightarrow\infty}F_{L=Na}(i\omega),
   \label{eq:sbbc}
\end{align}
with
\begin{align}
   w(i\omega)
   =&
   \frac{i}{4\pi}
   \int_{-\pi/a}^{\pi/a}\mathrm{d}k\:
   \mathrm{tr}
   \left[
      \Gamma G(i\omega,k)\partial_k G^{-1}(i\omega,k)
      \right],
   \label{eq:def_w}
   \\
   F_{L=Na}(i\omega)
   =&
   \sum_{l=1}^{2N}
   {\left[
         \gamma g(i\omega)
   \right]}_{l,l}
   \nonumber\\
   =&\sum_{j=1}^{N/2}[ g_{4j-3,4j-2}(i\omega)+g_{4j-2,4j-3}(i\omega)\nonumber\\
   &+g_{4j-2,4j}(i\omega)+g_{4j,4j-2}(i\omega)]
   \nonumber  \\
   =&
   \sum_{j=1}^{N/2}
   \left\{
      f_{\mathrm{AA}}^{\mathrm{intra}}(i\omega,j)
      +
      [f_{\mathrm{AA}}^{\mathrm{intra}}(i\omega,j)]^{\rm T}
      \right.
      \nonumber\\
   &
   \left.
      \hspace{8mm}+
      f_{\mathrm{BB}}^{\mathrm{intra}}(i\omega,j)
      +
     [ f_{\mathrm{BB}}^{\mathrm{intra}}(i\omega,j)]^{\rm T}
   \right\}
   \nonumber\\
   =&
   2i
   \sum_{j=1}^{N/2}
   \left[
      f_{\mathrm{AA}}^{\mathrm{OEE}}(i\omega,j)
      +
      f_{\mathrm{BB}}^{\mathrm{OEE}}(i\omega,j)
   \right],
   \label{eq:def_F}
\end{align}
where  $\Gamma=\sigma_0\tau_1$ is the chiral operator for the SC SSH model, $\gamma=\mathrm{diag}(\Gamma,\Gamma,\ldots)$,
$w(i\omega)$ is defined in a system with periodic boundary conditions, $F_{L=Na}$ is defined in a system with open boundary conditions, and $f_{\mathrm{AA(BB)}}^\mathrm{intra}$ and $f_{\mathrm{AA(BB)}}^\mathrm{OEE}$ are given by Eqs.~(\ref{eq23}) and (\ref{eq:FAAs}), respectively. In the last line of Eq.~(\ref{eq:def_F}), we use the fact that all matrix elements of the Hamiltonian in the real space basis is purely real and then the Green's function satisfies $g^{\rm T}=g$.
   Note that the components of the Green's function that are related to $F_{L=Na}(i\omega)$ depend on the chiral operator $\gamma$.
  Interestingly, the bulk value of
   $f_{\mathrm{AA}}^{\mathrm{OEE}}(i\omega,j)
   +
   f_{\mathrm{BB}}^{\mathrm{OEE}}(i\omega,j)$ is zero due to Eq.~(\ref{Pairamplitudes}) with $j=N/2$ is sufficiently far from both edges. This can be seen from the following:
   \begin{align}
      &
      \lim_{N\rightarrow\infty}
      \left[
         f_{\mathrm{AA}}^{\mathrm{OEE}}(i\omega,j=N/2)
         +
         f_{\mathrm{BB}}^{\mathrm{OEE}}(i\omega,j=N/2)
      \right]
      \nonumber\\
      =&
      \frac{a}{2\pi}
      \int_{-\pi/a}^{\pi/a}
      \mathrm{d}k\:
      \left[
         F_{\mathrm{AA}}(i\omega,k)
         +
         F_{\mathrm{BB}}(i\omega,k)
      \right]
      \nonumber\\
      =&
      \frac{a}{2\pi}
      \int_{-\pi/a}^{\pi/a}
      \mathrm{d}k\:
      \frac{1}{P(i\omega,k)}
      \left(
         \mu_\mathrm{B}
         +
         \mu_\mathrm{A}
      \right)
      \left(
         \Delta_k^*T_k
         -
         \Delta_k T_k^*
      \right)
      \nonumber\\
      =&0,
      \label{eq:sbbc_bulk_converge}
   \end{align}
since $\Delta_k^* T_k-\Delta_k T_k^*=4it\Delta(1-\eta\bar{\eta})\sin(ka)$ is an odd function of $k$. Thus, even though the individual terms $f^{\rm OEE}_{\rm AA}$ and $f^{\rm OEE}_{\rm BB}$  are non-zero in the bulk ($N/2$), the summation of them becomes zero as shown above.

By comparing Eqs.~(\ref{winding}) and (\ref{eq:def_w}), in the limit $\omega\rightarrow 0$, $w(i\omega)$ reduces to the winding number $W$, namely, $\lim_{\omega\rightarrow0}w(i\omega)=W$, where $W$ is given by Eq.\,(\ref{winding}).
In Eq.~(\ref{eq:def_F}), the summation runs from $j=1$ to $j=N/2$ in order  to extract the odd-$\omega$ amplitudes bounded close to the surface $j=1$. 
The reason why the summation is stopped at $N/2$ is that the signs of $f^{\rm OEE}_{\rm AA}+f^{\rm OEE}_{\rm BB}$ are opposite at the left and the right edges. A summation from $j=1$ to $N$ then vanishes. From the definition given by Eq.~(\ref{eq:def_F}), the total amount of intra-sublattice odd-$\omega$ pairing $f_\mathrm{AA(BB)}^\mathrm{OEE}$ accumulated close to the surface is connected to the generalized winding number $w(i\omega)$ through Eq.~(\ref{eq:sbbc}).

In order to provide further evidence of the relation between boundary odd-$\omega$ amplitudes and topological number,  in Fig.~\ref{fig:sbbc}(a,b)  we present numerical results of $w(i\omega)/(i\omega)$ and $F_{L=1000a}$.
We have verified (not shown) that the normalized difference between $w(i\omega)/(i\omega)$ and $F_{L=1000a}$  is smaller than $10^{-9}$. 
These figures demonstrate that $w(i\omega)/(i\omega)$ and $F_{L=1000a}$ are the same within numerical error.
In particular,  we can observe that $w(i\omega)/i\omega$ and $F_{L=1000a}(i\omega)$ have large values in the topological phase for small $\omega$.
This can be understood as follows. $F_{L=Na}$ is an odd function of $\omega$ and it can be expanded as
\begin{align}
   \lim_{N\rightarrow\infty}F_N(i\omega)
   =
   W/(i\omega)+i\chi \omega +\mathcal{O}({\omega}^3)
\end{align}
for small value of $\omega$, where $W$ is the winding number. 
From Eq.~(\ref{eq:sbbc}), $\chi$ is obtained from the second derivative of $w(i\omega)$ and we can confirm that
$\chi$ is a real number. From this expansion, the total amount of the odd-$\omega$ correlations $F_{L=Na}$ diverges in the topological phase ($W\neq0$) and it is a linear function of $\omega$ in the trivial phase ($W=0$) for small $\omega$. This is summarized  in Eq.\,(\ref{eq:sbbc_expansion}) of the main text.

\bibliography{biblio}

\end{document}